\documentclass[]{jfm}

\usepackage{graphicx}
\usepackage{newtxtext}
\usepackage{newtxmath}
\usepackage{natbib}
\usepackage{subfig}
 \usepackage[abs]{overpic}
\usepackage{placeins}
\usepackage{tikz}
\usepackage{hyperref}
\hypersetup{
    colorlinks = true,
    urlcolor   = blue,
    citecolor  = black,
}

\newcommand{\RomanNumeralCaps}[1]
\linenumbers

\title{Analysis of scale-dependent kinetic and potential energy in sheared, stably stratified turbulence}
\author{Xiaolong Zhang, Rohit Dhariwal, Gavin Portwood, Stephen M. de Bruyn Kops, Andrew D. Bragg}

\author{Xiaolong Zhang\aff{1},
  Rohit Dhariwal\aff{2},
  Gavin Portwood\aff{3},
  Stephen M. de Bruyn Kops \aff{4},
  Andrew D. Bragg\aff{1}
 \corresp{\email{andrew.bragg@duke.edu}}
 }

 \affiliation{\aff{1}Department of Civil and Environmental Engineering, Duke University, Durham, NC 27708, USA
 \aff{2}Center for Institutional Research Computing, Washington State University, Pullman, WA 99164, USA
  \aff{3}Lawrence Livermore National Laboratory, Livermore, CA 94550, USA
  \aff{4}Department of Mechanical and Industrial Engineering, University of Massachusetts Amherst, Amherst, MA 01003, USA}

\begin{document}
\maketitle

\begin{abstract}
Budgets of turbulent kinetic energy (TKE) and turbulent potential energy (TPE) at different scales $\ell$ in sheared, stably stratified turbulence are analyzed using a filtering approach. Competing effects in the flow are considered, along with the physical mechanisms governing the energy fluxes between scales, and the budgets are used to analyze data from direct numerical simulation (DNS) at buoyancy Reynolds number $Re_b=O(100)$. The mean TKE exceeds TPE by an order of magnitude at the large scales, with the difference reducing as $\ell$ is decreased. At larger scales, buoyancy is never observed to be positive, with buoyancy always converting TKE to TPE. As $\ell$ is decreased, the probability of locally convecting regions increases, though it remains small at scales down to the Ozmidov scale. The TKE and TPE fluxes between scales are both downscale on average and their instantaneous values are positively correlated, but not strongly so, and this occurs due to the different physical mechanisms that govern these fluxes. Moreover, the contribution to these fluxes arising from the sub-grid fields are shown to be significant, in addition to the filtered scale contributions associated with the processes of strain-self amplification, vortex stretching, and density gradient amplification. Probability density functions (PDFs) of the $Q,R$ invariants of the filtered velocity gradient are considered. Unlike isotropic turbulence, as $\ell$ increases the sheared-drop shape of the PDF disappears and the PDF becomes symmetric about $R=0$, meaning regions of vortex stretching and compression become equi-probable, as well as regions of strain amplification or suppression.
\end{abstract}

\section{Introduction}


When turbulence occurs in environmental flows, it is often affected by both mean-shear and stable stratification \citep{Vallis06,wyngaard,ferrari09,zorzetto18,ayet20}, leading to Sheared, Stably Stratified Turbulence (SSST). Mean-shear (vertical gradient of horizontal flow) produces turbulence, with hairpin vortices, streaks, and strong fluctuations in all three directions \citep{lee90,davidson,pope}. Stable stratification suppresses fluctuations in the vertical direction, and if sufficiently strong generates a quasi two-dimensional flow behavior \citep{riley00,riley12}. Stable stratification also provides a restoring buoyancy force that enables the propagation of internal waves and the formation of quasi-horizontal ``pancake'' vortical structures in the flow \citep{davidson}. Since mean-shear and stable stratification have competing effects, and produce different flow structures, the dynamics of SSST are rich and complex. Moreover, the flow is often ``patchy'', with turbulent and non-turbulent regions interspersed, depending on the local competition between shear and buoyancy. Indeed, when the local shear in the flow is strong enough for the local Richardson number to be sufficiently small, the local flow may undergo Kelvin-Helmholtz instabilities which can evolve due to nonlinearity into turbulent motion \citep{riley03}.

Understanding and modeling SSST is an active area of research with many open questions. One such issue concerns understanding the mixing efficiencies in SSST and their parametric dependence \citep{Peltier03,portwood19}, which are vital for predicting mixing in oceans \citep{Jayne09,Gregg18}. Another vital area is to understand the properties of SSST across its range of dynamical scales, and the physical mechanisms that govern the fluxes of turbulent kinetic and potential energy between scales. Not only is this important for a basic understanding of the flow, it is also of crucial importance for developing large eddy simulation (LES) models for SSST. A number of studies have considered the effect of stable stratification on the multiscale properties of turbulence using Fourier analysis and considering the average behavior of the flow in terms of the energy spectrum, the mean buoyancy and mean inter-scale energy transfer terms \citep{riley03,lindborg06,almalkie12}. The study of \cite{riley03} showed that the horizonal energy spectrum exhibits a $k_h^{-5/3}$ scaling for wavenumbers smaller than the Ozmidov wavenumber $k_O$ (corresponding to the wavenumber at which inertial and buoyancy forces are of the same order), where $k_h$ is the horizontal wavenumber, and the results indicated a downscale energy transfer of kinetic energy in the flow. This motivated \cite{lindborg06} who confirmed that strongly stratified turbulent flows exhibit a downscale cascade of turbulent kinetic and potential energy on average, and developed phenomenological predictions similar in spirit to Kolmogorov's 1941 theory \citep{kolmogorov41a}. The observation of a downscale energy cascade was contrary to predictions that had been made in \cite{gage79,lilly83} based on the assumption that in the limit of strong stratification, the flow should behave as two-dimensional turbulence. The basic reason why the predictions of \cite{gage79,lilly83} failed is that, as shown in \cite{billant01}, as the strength of the stratification increases and the vertical velocity of the flow is suppressed, the vertical lengthscale of the flow also reduces in such a way that the terms in the dynamical equations associated with vertical motion always remain $O(1)$, and hence the flow never becomes two-dimensional.

Many questions remain, however, regarding the multiscale properties of SSST. For example, how do fluctuations of the SSST flow about its mean-field state behave? What are the mechanisms of the turbulent kinetic and potential energy transfers among scales, and to what extent are these transfers coupled to each other, and to fluctuations in the local buoyancy? How do the sub-grid scale terms in SSST contribute to the flow dynamics across scales? To address these and other questions, in this paper we will use a filtering-based approach wherein the velocity and density fields are considered for different filtering lengths $\ell$. Using these we explore the turbulent kinetic and potential energies at different scales, and the various processes that determine their behavior. Using a filtering approach allows us to consider spatially local couplings between the processes that control the flow energetics at different scales, and to consider fluctuations in the flow as well as the mean-field behavior. Moreover, such an analysis of the multiscale properties of SSST can provide insights for developing large eddy simulation (LES) models of SSST, since LES models are often developed in the physical-space, rather than Fourier space context. LES modeling of stratified turbulence is an under-developed area \cite{khani15}, and new insights into the dynamics of stratified turbulence across scales could would aid in the development of appropriate sub-grid models.

The outline of the paper is as follows. In \S\ref{Theory}, we introduce the equations governing the turbulent kinetic and potential energy at different scales, consider the behavior of the mean-field state of the flow at different scales, and discuss the physical mechanisms governing the energy transfer among scales in SSST. In \S\ref{DNS} we describe the DNS dataset used in this study, and in \S\ref{RandD} we present and discuss results from the DNS for the quantities introduced in \S\ref{Theory} that govern the turbulent kinetic and potential energy at different scales. Finally, in \S\ref{Conc} we draw conclusions from the study and discuss future areas for investigation.

\section{Theoretical Considerations}\label{Theory}

\subsection{Governing equations for scale-dependent energy fields}

We consider the case where the mean velocity gradient, $\gamma$, and the mean density gradient, $\zeta$, are constant in space and time, such that the total fluid velocity vector can be written as $\boldsymbol{U}=z\gamma\boldsymbol{e}_x+\boldsymbol{u}$, where $\boldsymbol{u}$ is the fluctuating component of the velocity, and the total density as $\rho=\rho_r +z\zeta+\rho'$, where $\rho_r$ is a constant reference density and $\rho'$ is the fluctuating component of the density. While $\boldsymbol{e}_x$ is the unit vector in the direction of the mean velocity $\langle\boldsymbol{U}\rangle=z\gamma\boldsymbol{e}_x$, $\boldsymbol{e}_z$ is the unit vector pointing in the direction opposite to the gravitational acceleration $\boldsymbol{g}$ (i.e. the vertical direction), and $\boldsymbol{e}_y\equiv \boldsymbol{e}_z\times\boldsymbol{e}_x$.

The filtering operator to be used is defined for an arbitrary field $\boldsymbol{a}(\boldsymbol{x},t)$ as $\widetilde{\boldsymbol{a}}(\boldsymbol{x},t)\equiv\int_{\mathbb{R}^3}\boldsymbol{a}(\boldsymbol{x}-\boldsymbol{y},t)\mathcal{G}_\ell(\boldsymbol{y})\,d\boldsymbol{y}$, with $\mathcal{G}_\ell$ an isotropic filtering kernel with lengthscale $\ell$. In order for the energy fields introduced below to be strictly non-negative, which is a physical requirement, then a necessary and sufficient condition is that the filtering kernel $\mathcal{G}_\ell$ be non-negative \citep{vreman94}. We will use the isotropic Gaussian filter $\mathcal{G}_\ell(\boldsymbol{y})\equiv (2\pi\ell^2)^{-3/2}\exp(-\|\boldsymbol{y}\|^2/2\ell^2)$ which satisfies this condition.

Assuming that $|\rho'|/\rho\ll1$, the governing equation for $\boldsymbol{u}$ is then the Boussinesq-Navier-Stokes equation coupled with an advection-diffusion equation for $\rho'$ \citep{Vallis06}. The filtered versions of these equations are
\begin{align}
D_t\widetilde{\boldsymbol{u}}&=-(1/\rho_r)\boldsymbol{\nabla}\widetilde{p}-\boldsymbol{\nabla\cdot \tau}+2\nu\boldsymbol{\nabla\cdot}\widetilde{\boldsymbol{s}}- N\widetilde{\phi} \boldsymbol{e}_z+\widetilde{\boldsymbol{F}},\label{NSE_f}\\
D_t\widetilde{\phi}&=-\boldsymbol{\nabla\cdot \Sigma}+\kappa\nabla^2\widetilde{\phi} +N \widetilde{u}_z+\widetilde{f},\label{rho_f}
\end{align}
where ${D_t\equiv \partial_t+\widetilde{\boldsymbol{u}}\boldsymbol{\cdot\nabla}}$, $\widetilde{\phi}\equiv g\widetilde{\rho'}/N\rho_r$ is the scaled density field (with dimensions of a velocity), $g\equiv\|\boldsymbol{g}\|$, $\nu$ and $\kappa$ are the kinematic viscosity and thermal diffusivity, respectively, $N\equiv\sqrt{-g\zeta/\rho_r}$ is the buoyancy frequency, $\boldsymbol{\tau}\equiv\widetilde{\boldsymbol{uu}}-\widetilde{\boldsymbol{u}}\widetilde{\boldsymbol{u}}$ is the sub-grid stress tensor, $\widetilde{\boldsymbol{s}}\equiv (\boldsymbol{\nabla}\widetilde{\boldsymbol{u}}+[\boldsymbol{\nabla}\widetilde{\boldsymbol{u}}]^\top)/2$ is the filtered strain-rate, and $\boldsymbol{\Sigma}\equiv\widetilde{\boldsymbol{u}\phi}-\widetilde{\boldsymbol{u}}\widetilde{\phi}$ is the sub-grid vector for the scaled density field.  The terms $\boldsymbol{F}$ and $f$, whose filtered forms appear in the equations above, are the forcing terms associated with mean velocity gradient applied to the flow, and for the SSST under consideration are given by
\begin{align}
\boldsymbol{F}&=-\gamma(\boldsymbol{e}_z\boldsymbol{\cdot x})(\boldsymbol{e}_x \boldsymbol{\cdot \nabla})\boldsymbol{u} -\gamma\boldsymbol{e}_x(\boldsymbol{e}_z\boldsymbol{\cdot u}),\\
f&=-\gamma(\boldsymbol{e}_z\boldsymbol{\cdot x})(\boldsymbol{e}_x \boldsymbol{\cdot \nabla})\phi.
\end{align}

Following \cite{germano92}, the turbulent kinetic energy (TKE) in a given region of size $\ell$ is $\widetilde{\boldsymbol{u\cdot u}}/2$ and this may be split up into the TKE at scales $\geq \ell$ (``large-scales'') denoted by $E_K\equiv \|\widetilde{\boldsymbol{u}}\|^2/2$ and the TKE at scales $< \ell$ (``small-scales'') denoted by $e_K\equiv (\widetilde{\|\boldsymbol{u}\|^2}-\|\widetilde{\boldsymbol{u}}\|^2)/2=\mathrm{tr}[\boldsymbol{\tau}]/2$. The equations governing $E_K$ and $e_K$ may be derived from \eqref{NSE_f} and are given by
\begin{align}
D_t E_K&=-(1/\rho_r)\boldsymbol{\nabla\cdot}(\widetilde{\boldsymbol{u}}\widetilde{p})-\boldsymbol{\nabla\cdot}(\widetilde{\boldsymbol{u}}\boldsymbol{\cdot \tau})-\Pi_K
+2\nu\boldsymbol{\nabla\cdot}(\widetilde{\boldsymbol{u}}\boldsymbol{\cdot} \widetilde{\boldsymbol{s}})-2\nu\|\widetilde{\boldsymbol{s}}\|^2
-N\widetilde{{u}}_z\widetilde{\phi}+\widetilde{\boldsymbol{F}}\boldsymbol{\cdot} \widetilde{\boldsymbol{ u}},\\
\begin{split}
D_t e_K&=\boldsymbol{\nabla\cdot }\boldsymbol{T}_K +\mathcal{B}+\Pi_K-\varepsilon_K+\mathcal{F}_K.\label{SS_TKE}
\end{split}
\end{align}
In these equations, the scale-to-scale TKE flux is defined as ${\Pi_K\equiv-\boldsymbol{\tau:}\widetilde{\boldsymbol{s}}}$, such that $\Pi_K>0$ corresponds to a transfer of TKE from the large to the small scales. The small-scale buoyancy term is $\mathcal{B}\equiv -N(\widetilde{u_z\phi}-\widetilde{{u}}_z\widetilde{\phi})$, the small-scale TKE dissipation-rate is $\varepsilon_K\equiv 2\nu(\widetilde{\|\boldsymbol{s}\|^2}-\|\widetilde{\boldsymbol{s}}\|^2)$, the small-scale forcing is $\mathcal{F}_K\equiv \widetilde{\boldsymbol{F\cdot u}}-\widetilde{\boldsymbol{F}}\boldsymbol{\cdot}\widetilde{\boldsymbol{u}}$, and the small-scale TKE transport term involves
\[\boldsymbol{T}_K\equiv -(1/\rho_r)(\widetilde{\boldsymbol{u}p} -\widetilde{\boldsymbol{u}}\widetilde{p})-(1/2)(\widetilde{\boldsymbol{u} \|\boldsymbol{u}\|^2}-\widetilde{\boldsymbol{u}}\widetilde{\|\boldsymbol{u}\|^2})+(\widetilde{\boldsymbol{u}}\boldsymbol{\cdot \tau})+2\nu(\widetilde{\boldsymbol{u\cdot s}} -\widetilde{\boldsymbol{u}}\boldsymbol{\cdot} \widetilde{\boldsymbol{s}}).\]
The turbulent potential energy (TPE) in a given region of size $\ell$ is $\widetilde{\phi\phi}$, and this may be split up into the amount contained in the large scales denoted by $E_P\equiv\widetilde{\phi}\widetilde{\phi}/2$ and the amount contained in the small-scales denoted by $e_P\equiv(\widetilde{\phi\phi}-\widetilde{\phi}\widetilde{\phi})/2$. The equations governing $E_P$ and $e_P$ may be derived from \eqref{rho_f} and are given by
\begin{align}
D_t E_P&=-\boldsymbol{\nabla\cdot}(\widetilde{\phi}\boldsymbol{\Sigma})-\Pi_P   +\frac{1}{2}\kappa\nabla^2\widetilde{\phi}\widetilde{\phi}-\kappa\|\boldsymbol{\nabla}\widetilde{\phi}\|^2 +N \widetilde{\phi}\widetilde{u}_z+\widetilde{\phi}\widetilde{f},\\
\begin{split}
D_t e_P&=\boldsymbol{\nabla\cdot}\boldsymbol{T}_P-\varepsilon_P-\mathcal{B}+\Pi_P+\mathcal{F}_P.\label{SS_TPE}
\end{split}
\end{align}
In these equations, the scale-to-scale TPE flux is defined as $\Pi_P\equiv -\boldsymbol{\Sigma\cdot\nabla}\widetilde{\phi}$, such that $\Pi_P>0$ corresponds to a transfer of TPE from the large to the small scales. The small-scale TPE dissipation-rate is $\varepsilon_P\equiv \kappa(\widetilde{\|\boldsymbol{\nabla}\phi\|^2}-\|\boldsymbol{\nabla}\widetilde{\phi}\|^2)$, the small-scale scalar forcing is $\mathcal{F}_P\equiv \widetilde{\phi f}-\widetilde{\phi}\widetilde{f}$, and the small-scale scalar transport involves \[\boldsymbol{T}_P\equiv (\kappa/2)\boldsymbol{\nabla}(\widetilde{\phi\phi}-\widetilde{\phi}\widetilde{\phi})+\widetilde{\phi}\boldsymbol{\Sigma}.\]

In the following, attention will be given to the behavior of the terms that contribute to $D_t e_K$ and $D_t e_P$. Note that while $e_K$ and $e_P$ are referred to as the small-scale TKE and TPE, respectively, they correspond to the TKE and TPE contained in all scales $<\ell$, so that when $\ell$ exceeds the integral legnthscale of the flow, $e_K$ and $e_P$ actually contain the contributions from the largest dynamical scales in the flow.

\subsection{Length scales in SSST}\label{scales}

As discussed in \cite{portwood19}, for the velocity field there are four important length scales for the SSST flow under consideration. The first is the large-eddy lengthscale which may be characterized by $L=\mathcal{E}_K^{3/2}/\langle\epsilon_K\rangle$, where $\mathcal{E}_K\equiv\lim_{\ell\to\infty}\langle e_K\rangle$ is the total (i.e. involving contributions from all scales) mean TKE in the flow, and $\langle\epsilon_K\rangle\equiv\lim_{\ell\to\infty}\langle \varepsilon_K\rangle$ is the total mean TKE dissipation-rate. The second is the Kolmogorov lengthscale $\eta=(\nu^3/\langle\epsilon_K\rangle)^{1/4}$ which characterizes the scale at which viscous and inertial scales in the flow are of the same order. The third is the Ozmidov lengthscale $\ell_O=(\langle \epsilon_K\rangle/N^3)^{1/2}$, which characterizes the scale at which buoyancy and inertial forces are of the same order. The fourth is the Corrsin lengthscale $\ell_C=(\langle\epsilon_K\rangle/\gamma^3)^{1/2}$, which characterizes the scale at which mean-shear and inertial forces are of the same order. Different nondimensional flow parameters may be related to these lengthcales (assuming $\nu=\kappa$)
\begin{align}
Ri&\equiv\frac{N^2}{\gamma^2}=\Big(\ell_C/\ell_O\Big)^{4/3},\\
Fr&\equiv\frac{\langle\epsilon_K\rangle}{N\mathcal{E}_K}=\Big(\ell_O/L\Big)^{2/3},\\
Re_b&\equiv\frac{\langle\epsilon_K\rangle}{\nu N^2}=\Big(\ell_O/\eta\Big)^{4/3},\\
Re_s&\equiv\frac{\langle\epsilon_K\rangle}{\nu \gamma^2}=\Big(\ell_C/\eta\Big)^{4/3}.
\end{align}
In order for the flow to become turbulent, the Richardson number must satisfy $Ri<O(1)$ (and therefore $\ell_C<O(\ell_O)$), and in order for stratification to have an impact on the flow the Froude number must satisfy $Fr<O(1)$. In order for there is to exist a range scales $\ell\ll\ell_O$ that are not affected by buoyancy, the buoyancy Reynolds number must satisfy $Re_b\gg 1$. Correspondingly, in order for there is to exist a range scales $\ell\ll\ell_C$ that are not affected by the mean-shear  the shear Reynolds number must satisfy $Re_s\gg 1$. These Reynolds numbers are, however, related since $Re_s=Re_b Ri$. 

For the scalar field, the large scale is $L_P\equiv \mathcal{E}_P^{3/2}\langle\epsilon_K\rangle^{1/2}/\langle\epsilon_P\rangle^{3/2}$, where $\mathcal{E}_P\equiv\lim_{\ell\to\infty}\langle e_P\rangle$ is the total mean TPE in the flow, and $\langle\epsilon_P\rangle\equiv\lim_{\ell\to\infty}\langle \varepsilon_P\rangle$ is the total mean TPE dissipation-rate. The small scale is the Batchelor scale $\eta_B\equiv (\nu\kappa^2/\langle\epsilon_K\rangle)^{1/4}$. In this paper we focus on flows with $\nu=\kappa$ such that $\eta_B=\eta$.

\subsection{Mean-field behavior}

We now turn to consider the contributions to the equations governing the mean energy fields $\langle e_K\rangle$ and $\langle e_P\rangle$. For SSST, the mean transport equations for these quantities reduce to
\begin{align}
0&=\langle\mathcal{B}\rangle-\langle\varepsilon_K\rangle+\langle\Pi_K\rangle+\langle\mathcal{F}_K\rangle,\label{TKE_budget_2}\\
0&=-\langle\mathcal{B}\rangle-\langle\varepsilon_P\rangle+\langle\Pi_P\rangle.\label{PE_budget_2}
\end{align}
In the following discussion, we will first consider the case where the forcing is confined to the large scales (such as in \cite{lindborg06}), a flow we here refer to as forced stably stratified turbulence (FSST), for which we introduce the scale $\ell_F$ as the scale below which the forcing plays a sub-leading role in in the flow. This behavior will then be compared to that of SSST to understand the differences. 


Focussing first on the TKE, in the limit $\ell/\eta \to\infty$, $\langle\Pi_K\rangle\to0$, $\langle\varepsilon_K\rangle\to\langle\epsilon_K\rangle$, $\langle\mathcal{F}_K\rangle\to\langle\mathcal{F}_K\rangle^{\infty}$, $\langle\mathcal{B}\rangle\to\langle\mathcal{B}\rangle^\infty$ , so that we have
\begin{align}
\langle\mathcal{F}_K\rangle^\infty&\sim-\langle\mathcal{B}\rangle^\infty+\langle\epsilon_K\rangle,
\end{align}
reflecting a balance between the total injection of TKE by the forcing and the total energy lost due to viscous dissipation and conversion to TPE. The behavior of \eqref{TKE_budget_2} and \eqref{PE_budget_2} as the scale $\ell$ decreases depends on the dynamical scales of the system that were introduced in \S\ref{scales}. 

For FSST $\langle\Pi_K\rangle\sim-\langle\mathcal{B}\rangle+\langle\epsilon_K\rangle$ when $\ell_F\gg\ell\gg\eta$, and for $\ell_O/\ell_F\to \infty$ (neutrally buoyant), this would correspond to a TKE cascade $\langle\Pi_K\rangle\sim\langle\epsilon_K\rangle$. However, for a stably stratified flow where $\langle\mathcal{B}\rangle^{\infty}<0$, since buoyancy effects reduce with decreasing scale, $\langle\Pi_K\rangle\sim-\langle\mathcal{B}\rangle+\langle\epsilon_K\rangle$ implies that the TKE flux $\langle\Pi_K\rangle$ will actually reduce as $\ell$ decreases, until it approaches a constant value in the regime $\ell_O>\ell\gg\eta$, as discussed in \cite{riley12,kumar14}. As a result of this, the TKE flux in stratified turbulence cannot be in the form of a cascade (which would require a constant energy flux). This is simply a reflection of the fact that TKE is being lost as it is passed down to smaller scales due to conversion of TKE to TPE.

For SSST, the behavior of $\langle\Pi_K\rangle$ is quite different. In this case, since $\langle\mathcal{F}_K\rangle$ operates down to the scale $\ell_C$, then because $\ell_C<\ell_O$ the behavior $\langle\Pi_K\rangle\sim-\langle\mathcal{B}\rangle+\langle\epsilon_K\rangle$ never emerges, and the regime that emerges instead for $Ri\ll 1$ is $\langle\Pi_K\rangle\sim-\langle\mathcal{F}_K\rangle+\langle\epsilon_K\rangle$ for $\ell\gg\eta$. In this regime, the TKE flux is again not in the form of a cascade since $\langle\mathcal{F}_K\rangle$ depends on $\ell$. Indeed, based on the behavior of the co-spectrum \citep{katul13} we might expect the following behavior to emerge at high $Re$ in SSST
\begin{align}
\langle\Pi_K\rangle\sim -\gamma^2\langle\epsilon_K\rangle^{1/3}\ell^{4/3}+\langle\epsilon_K\rangle,\quad\text{for}\quad L\gg \ell\gg\eta.
\end{align}
Therefore, in this range $\langle\Pi_K\rangle$ increases as $\ell$ decreases, until it asymptotes to the constant flux cascade regime $\langle\Pi_K\rangle\sim \langle\epsilon_K\rangle$ for $\ell_C\gg\ell\gg\eta$. This is the opposite behavior to that discussed in \cite{kumar14} for FSST where the forcing is confined to scales $\gg\ell_O$ and where $\langle\Pi_K\rangle$ reduces as $\ell$ decreases towards $\ell_O$.

In the above discussion, it has been assumed that $\langle\varepsilon_K\rangle\sim\langle\epsilon_K\rangle$ for $\ell\gg \eta$. While this is true for isotropic turbulence, it will not in general apply in strongly stratified flows with $Fr\ll 1$. This is because when $Fr\ll1$, the flow structures are highly anisotropic, and so while the horizontal lengthscale may be large enough for viscous effects to be negligible, the vertical lengthscale of the structure may be small enough for viscous effects to be important \citep{riley12}. In such a situation, even if $\langle\Pi_K\rangle\sim\langle\varepsilon_K\rangle$ emerges in some range of scales, this would not correspond to an inertial cascade since $\langle\varepsilon_K\rangle$ would depend on $\ell$.

Concerning the TPE field, for flows with $\nu=\kappa$, then for $\ell\gg\eta$ 
\begin{align}
\langle\Pi_P\rangle\sim\langle\mathcal{B}\rangle+\langle\epsilon_P\rangle.
\end{align}
Hence, in the regime $\ell>\ell_O$, as $\ell$ decreases, $\langle\Pi_P\rangle$ increases, while for $\ell_O\gg\ell\gg\eta$, the constant flux cascade regime $\langle\Pi_P\rangle\sim\langle\epsilon_P\rangle$ emerges. Therefore, there cannot be a TPE cascade in the strict sense at scales where density is an active scalar, i.e. $\ell>\ell_O$, since the scale-dependency of $\langle\mathcal{B}\rangle$ at these scales leads to a non-constant flux of TPE at these scales. This observation seems to be in conflict with the well-known Bolgiano-Obhukov (BO) scaling \citep{bolgiano59,obukhov59} that has been proposed for stratified turbulent flows, in which $\langle\Pi_K\rangle$ is supposed to decay with reducing $\ell$ due to buoyancy as $\langle\Pi_K\rangle\propto \ell^{4/5}$ while $\langle\Pi_P\rangle$ is constant $\langle\Pi_P\rangle\sim \langle\epsilon_P\rangle$. This behavior, however, seems problematic. For example, in FSST then for $\ell_F\gg \ell\gg\eta$ 
\begin{align}
\langle\Pi_K\rangle&\sim-\langle\mathcal{B}\rangle+\langle\epsilon_K\rangle,\\
\langle\Pi_P\rangle&\sim\langle\mathcal{B}\rangle+\langle\epsilon_P\rangle.
\end{align}
It would seem that BO scaling could only emerge if $\langle\epsilon_P\rangle\gg \langle\epsilon_K\rangle=O(|\langle\mathcal{B}\rangle| )$, as noted in \cite{alam19}, so that we could have $\langle\Pi_K\rangle\sim-\langle\mathcal{B}\rangle+\langle\epsilon_K\rangle$ in which $\langle\Pi_K\rangle$ reduces as $\ell$ reduces while $\langle\Pi_P\rangle\sim\langle\epsilon_P\rangle$. However, DNS of FSST show $\langle\epsilon_P\rangle\lesssim \langle\epsilon_K\rangle$ over a range of $Fr$ and $Re_b$ \citep{lindborg06,debk15} which includes the ``moderately stratified'' regime $Re\gg1$ and $Fr\approx 1$ for which it has been argued that BO should apply \citep{alam19}. In SSST, BO scaling cannot emerge because there are by definition no scales at which the dynamics are given by a balance between inertia and buoyancy forces, since the mean-shear production is larger than the buoyancy at all scales at which it is active, reflected in that SSST requires $\ell_C<\ell_O$.

\subsection{Mechanisms governing the TKE and TPE fluxes}\label{FMech}

Since stable stratification acts to two-dimensionalize the flow, it had been conjectured that such flows might feature an inverse energy cascade \citep{gage79,lilly83}. However, it has now been demonstrated that this does not occur, but that there is a forward/downscale cascade of both kinetic and potential energy \citep{riley03,waite06,lindborg05,lindborg06,brethouwer07}. The TKE cascade in homogeneous, isotropic turbulence (HIT) is also downscale on average in three-dimensions, however, the underlying mechanisms driving the energy transfers in HIT and SSST might be quite different. In the context of HIT, it has long been thought that the key mechanism driving the TKE cascade is that of vortex stretching (VS) \citep{taylor22,taylor38,tennekes,davidson,doan18}. However, recent studies have demonstrated quantitatively that while VS plays an important role, the largest contribution to the TKE cascade comes from the dynamical process of the self-amplification of the strain-rate field \citep{carbone20,johnson20,johnson21}. An important question is how this understanding applies to SSST, where effects such as internal waves and mean-shear can play a role (whether directly or indirectly) in how TKE is transferred between scales, as well as the question of the mechanism driving the TPE transfer.

In \cite{johnson20,johnson21} a powerful, exact relationship was derived for $\Pi_K$ that assumes only that the filtering kernel $\mathcal{G}_\ell$ used in constructing $\widetilde{\boldsymbol{u}}$ is Gaussian. The result is
\begin{align}
\Pi_K&=\Pi_K^{F,SSA}+\Pi_K^{F,VS}+\Pi_K^{SG,SSA}+\Pi_K^{SG,VS}+\Pi_K^{SG,C},\label{Pik_sol}\\
\Pi_K^{F,SSA}&=-\ell^2\boldsymbol{\widetilde{s}}\boldsymbol{:}(\boldsymbol{\widetilde{s}}\boldsymbol{ \cdot}\boldsymbol{\widetilde{s}}),\\
\Pi_K^{F,VS}&=(1/4)\ell^2 \boldsymbol{\widetilde{s}} \boldsymbol{:}(\boldsymbol{\widetilde{\omega}}\boldsymbol{\widetilde{\omega}}),
\end{align}
where superscript `$F$' denotes that the quantity only depends on the filtered fields, while superscript `$SG$' denotes that the quantity depends on the sub-grid fields as well as the filtered fields. Explicit integral formalae for the $SG$ terms can be found in \cite{johnson20,johnson21}; we do not quote them here as they will not be considered in detail in the present paper. Note that in \cite{johnson20,johnson21}, the  `$F$' contributions are referred to as the ``local'' contributions while the  `$SG$' contributions are referred to as ``non-local'' contributions. While the use of this terminology is appropriate in studies of turbulence \citep{lvov92,aluie09}, we prefer to avoid it, because in turbulence, a local energy flux is usually typically taken to imply an energy flux dominated by the interaction between scales of similar size, which is not necessarily a property possessed by $\Pi_K^{F,SSA}$ and $\Pi_K^{F,VS}$ which in principle may involve the interaction of any scales in the flow of size $\geq \ell$.

The term $\Pi_K^{F,SSA}$ describes the contribution to the TKE flux arising from the process of strain self-amplification (SSA) whereby the filtered strain-rate field $\boldsymbol{\widetilde{s}}$ interacts with itself due to nonlinearity to either amplify (if $\boldsymbol{\widetilde{s}}\boldsymbol{:}(\boldsymbol{\widetilde{s}}\boldsymbol{ \cdot}\boldsymbol{\widetilde{s}})<0$) or else suppress (if $\boldsymbol{\widetilde{s}}\boldsymbol{:}(\boldsymbol{\widetilde{s}}\boldsymbol{ \cdot}\boldsymbol{\widetilde{s}})>0$) the strain-rate magnitude $\|\boldsymbol{\widetilde{s}}\|$. The term $\Pi_K^{F,VS}$ describes the contribution to the TKE flux arising from the process of vortex stretching (VS) whereby the filtered strain-rate field $\boldsymbol{\widetilde{s}}$ either amplifies (if $\boldsymbol{\widetilde{s}} \boldsymbol{:}(\boldsymbol{\widetilde{\omega}}\boldsymbol{\widetilde{\omega}})>0$) or else suppresses (if $\boldsymbol{\widetilde{s}} \boldsymbol{:}(\boldsymbol{\widetilde{\omega}}\boldsymbol{\widetilde{\omega}})<0$) the enstrophy $\|\boldsymbol{\widetilde{\omega}}\|^2$ through the process of vortex stretching (or compression).

The average of the contribution from the filtered field $\Pi_K^{F}=\Pi_K^{F,SSA}+\Pi_K^{F,VS}$ in \eqref{Pik_sol} is similar to the expression for the TKE flux in a two-point Karman-Hawarth equation derived using filtering and an asymptotic expansion \citep{carbone20}. Moreover, due to the relation of \cite{betchov56}, $\langle \Pi_K^{F,SSA}\rangle=3\langle \Pi_K^{F,VS}\rangle$ for incompressible, homogeneous turbulence, so that the contribution of SSA to $\langle \Pi_K^{F}\rangle$ is three times larger than that from VS. The sub-grid contributions $\Pi_K^{SG,SSA}$ and $\Pi_K^{SG,VS}$ are similar to $\Pi_K^{F,SSA}$ and $\Pi_K^{F,VS}$ except that in the $SG$ contributions, the filtered strain-rate $\boldsymbol{\widetilde{s}}$ acts to amply the strain-rate at sub-grid scales (for  $\Pi_K^{SG,SSA}$) and vorticity at sub-grid scales (for  $\Pi_K^{SG,VS}$). DNS data in \cite{johnson20,johnson21} showed that $\langle\Pi_K^{SG,SSA}\rangle\approx\langle\Pi_K^{SG,VS}\rangle$ in the inertial range of isotropic turbulence, so that overall, the strain-self amplification mechanism contributes more to the average energy cascade $\langle \Pi_K\rangle$ than vortex stretching does, contrary to the traditional explanation according to which vortex stretching is seen as the key mechanism driving the energy cascade \citep{taylor38,tennekes,davidson,doan18}.

The result in \eqref{Pik_sol} also applies to SSST because it makes no assumption about the flow dynamics. As such, in SSST, SSA and VS will still be the key dynamical processes governing the TKE flux. Moreover, the relation of \cite{betchov56} still applies (since it only assumes incompressibility and homogeneity; it does not assume isotropy). Therefore, at least with respect to the filtered field contribution $\langle \Pi_K^{F}\rangle$, SSA still contributes three times as much as VS to the total TKE flux. The relative contribution of $\langle \Pi_K^{SG,SSA}\rangle$ and $\langle \Pi_K^{SG,VS}\rangle$ to $\langle \Pi_K^{SG}\rangle$, however, may differ from that in HIT. Moreover, the actual behavior of the SSA and VS processes in SSST may differ appreciably from that in isotropic turbulence, since, for example, in SSST internal waves can contribute to the behavior of $\boldsymbol{\widetilde{s}}$ and $\boldsymbol{\widetilde{\omega}}$, and anisotropy in the flow can modify the alignments between $\boldsymbol{\widetilde{s}}$ and $\boldsymbol{\widetilde{\omega}}$ (see \cite{Sujovolsky20}) which affects the VS process. Some of these more involved questions will be the subject of a forthcoming work. A key point to be explored here is the relative contributions of $\Pi_K^{F}$ and $\Pi_K^{SG}$ to $\Pi_K$, and the correlations between these terms which is important to understand for LES modeling of SSST.

Following the same procedure outlined in \cite{johnson20,johnson21}, a result analogous to \eqref{Pik_sol} can be derived for $\Pi_P$
\begin{align}
\Pi_P&=\Pi_P^{F}+\Pi_P^{SG,S}+\Pi_P^{SG,V},\label{PiP_sol}\\
\Pi_P^{F}&=-\ell^2\boldsymbol{\widetilde{s}}\boldsymbol{:}(\boldsymbol{\nabla}\widetilde{\phi}\boldsymbol{\nabla}\widetilde{\phi}),\\
\Pi_P^{SG,S}&=-\boldsymbol{\nabla}\widetilde{\phi}\boldsymbol{\cdot}\int_0^{\ell^2} \tau_\beta\Big(\boldsymbol{\nabla}\widetilde{\phi}^{\sqrt{\alpha}},\boldsymbol{\widetilde{s}}^{\sqrt{\alpha}}\Big)\,d\alpha,\\
\Pi_P^{SG,V}&=-\boldsymbol{\nabla}\widetilde{\phi}\boldsymbol{\cdot}\int_0^{\ell^2} \tau_\beta\Big(\boldsymbol{\nabla}\widetilde{\phi}^{\sqrt{\alpha}},\boldsymbol{\widetilde{r}}^{\sqrt{\alpha}}\Big)\,d\alpha,
\end{align}
where $\widetilde{(\cdot)}^{\sqrt{\alpha}}$ denotes filtering at scale $\sqrt{\alpha}$ (rather than scale $\ell$), and for arbitrary first order $\boldsymbol{p}$ and second order $\boldsymbol{q}$ tensor fields, $\tau_\beta$ is defined as $\tau_\beta(\boldsymbol{p},\boldsymbol{q})\equiv\widetilde{\boldsymbol{p\cdot q}}^{\beta}-\widetilde{\boldsymbol{p}}^{\beta}\boldsymbol{\cdot}\widetilde{\boldsymbol{q}}^{\beta}$ with $\beta\equiv\sqrt{\ell^2-\alpha}$, and ${\boldsymbol{r}}\equiv (\boldsymbol{\nabla}{\boldsymbol{u}}-[\boldsymbol{\nabla}{\boldsymbol{u}}]^\top)/2$ is the rotation-rate tensor. The rotational motion in the flow makes no explicit contribution to the filtered flux $\Pi_P^{F}$ because $\boldsymbol{\widetilde{r}}\boldsymbol{:}(\boldsymbol{\nabla}\widetilde{\phi}\boldsymbol{\nabla}\widetilde{\phi})=0$, although it does affect $\Pi_P^{F}$ implicitly since rotation in the fluid affects the alignment between $\boldsymbol{\widetilde{s}}$ and $\boldsymbol{\nabla}\widetilde{\phi}$.

The contribution from the filtered field $\Pi_P^{F}$ describes the flux of TPE associated with the amplification (if $\boldsymbol{\widetilde{s}}\boldsymbol{:}(\boldsymbol{\nabla}\widetilde{\phi}\boldsymbol{\nabla}\widetilde{\phi})<0$) or suppression (if $\boldsymbol{\widetilde{s}}\boldsymbol{:}(\boldsymbol{\nabla}\widetilde{\phi}\boldsymbol{\nabla}\widetilde{\phi})>0$) of $\|\boldsymbol{\nabla}\widetilde{\phi}\|$ due to the filtered strain-rate $\boldsymbol{\widetilde{s}}$. The contribution $\Pi_P^{F}$ is similar in form to $\Pi_K^{F,VS}$, in that both depend upon the strain-rate field $\boldsymbol{\widetilde{s}}$ acting to amplify or suppress another dynamical field, rather than SSA in which the strain-rate field amplifies or suppresses itself. However, the negative sign appearing in $\Pi_P^{F}$ but not in $\Pi_K^{F,VS}$ leads to a significant difference in how the strain-rate contributes to these energy fluxes. In particular, if we define $\widetilde{\lambda}_i$ and $\widetilde{\boldsymbol{v}}_i$ as the $i^{th}$ ordered eigenvalues and eigenvectors of $\boldsymbol{\widetilde{s}}$, then $\langle\Pi_K^{F,VS}\rangle= (1/4)\ell^2\langle\widetilde{\lambda}_i\|\boldsymbol{\widetilde{\omega}}\|^2 \cos^2\theta_{\omega,i}\rangle$ and $\langle\Pi_P^{F}\rangle= -\ell^2\langle\widetilde{\lambda}_i\|\boldsymbol{\nabla\widetilde{\phi}}\|^2 \cos^2\theta_{\nabla\phi ,i}\rangle$, where $\theta_{\omega,i}$ denotes the angle between $\boldsymbol{\widetilde{\omega}}$ and $\boldsymbol{v}_i$, and similarly for $\theta_{\nabla\phi ,i}$. In view of this, if $\langle\Pi_K^{F,VS}\rangle>0$ so that VS contributes to the average downscale TKE flux, then $\langle\widetilde{\lambda}_i\|\boldsymbol{\widetilde{\omega}}\|^2 \cos^2\theta_{\omega,i}\rangle$ must be dominated by contributions from the extensional eigendirections of $\boldsymbol{\widetilde{s}}$. In isotropic turbulence and for $\ell/\eta\to 0$, $\boldsymbol{\widetilde{\omega}}$ preferentially aligns with $\boldsymbol{v}_2$, but the contribution to $\langle\widetilde{\lambda}_i\|\boldsymbol{\widetilde{\omega}}\|^2 \cos^2\theta_{\omega,i}\rangle$ from $\widetilde{\lambda}_1$ is larger than that from $\widetilde{\lambda}_2$ since $\widetilde{\lambda}_1$ tends to be much larger than the positive values of $\widetilde{\lambda}_2$ \citep{tsinober,buaria20}. This is all the more true for $\ell/\eta$ in the inertial range where the contribution from $\widetilde{\lambda}_1$ to $\langle\widetilde{\lambda}_i\|\boldsymbol{\widetilde{\omega}}\|^2 \cos^2\theta_{\omega,i}\rangle$ is much larger than that from $\widetilde{\lambda}_2$ \citep{carbone20}. In an analogous way, if $\langle\Pi_P^{F}\rangle= -\ell^2\langle\widetilde{\lambda}_i\|\boldsymbol{\nabla\widetilde{\phi}}\|^2 \cos^2\theta_{\nabla\phi ,i}\rangle>0$, then $\langle\widetilde{\lambda}_i\|\boldsymbol{\nabla\widetilde{\phi}}\|^2 \cos^2\theta_{\nabla\phi ,i}\rangle$ must be dominated by contributions from the compressive eigendirections of $\boldsymbol{\widetilde{s}}$. Hence, while the VS contribution to the downscale TKE flux is generated by the amplification of $\boldsymbol{\widetilde{\omega}}$ due to the extensional straining motions in the flow, the downscale TPE flux is generated by the amplification of $\boldsymbol{\nabla}\widetilde{\phi}$ due to compressional straining motions in the flow.

The sub-grid TPE flux associated with the strain-rate field $\Pi_P^{SG,S}$ and rotation-rate field $\Pi_P^{SG,V}$ do not have simple interpretations (unlike $\Pi_K^{SG,SSA}$ and $\Pi_K^{SG,VS}$), but are related to the amplification of the scalar gradients at multiple scales. It is interesting, however, that the rotation-rate (and therefore vorticity) field makes an explicit contribution to the TPE flux through the sub-grid term $\Pi_P^{SG,V}$, even though it makes no explicit contribution to the filtered flux $\Pi_P^{F}$.

The discussion above highlights that the physical mechanisms governing the TKE and TPE fluxes in stratified turbulence are quite different, so that while both are positive on average (e.g. \cite{lindborg06}), it may not be reasonable to model them in similar ways. We will explore this further by considering the statistical correlation between $\Pi_K$ and $\Pi_P$ which will provide insights into the extent to which they might be modeled using similar sub-grid closures in the context of LES.



\section{Direct Numerical Simulations}\label{DNS}

\begin{table}
\caption{Table of parameters in DNS.} 
\centering 
\begin{tabular}{c c c c c c} 
\hline\hline 
$Re_b$ & $Ri$ & $Fr$ & $L/\eta$ & $\ell_O/\eta$ & $\ell_C/\eta$ \\ [0.5ex] 
\hline 
81 & 0.154 & 0.50 & 76.37 & 27 & 6.64 \\ [1ex] 
\hline 
\end{tabular}
\label{table:DNS} 
\end{table}
In the next section, data from DNS will be analyzed by computing various terms in \eqref{SS_TKE} and \eqref{SS_TPE} to understand the processes and mechanisms controlling the behavior of the TKE and TPE across scales in SSST. The DNS data used is from the same data set presented in \cite{portwood19}, which we here summarize. In the DNS, the unfiltered versions of \eqref{NSE_f} and \eqref{rho_f} are solved with constant mean velocity gradient $\gamma$ and mean density gradient $\zeta$ using the Fourier pseudospectral scheme described in \cite{debk15,almalkie12} but with the mean-shear term handled using an integrating factor \citep{brucker07,chung12,sekimoto16}.

In order to generate a statistically stationary flow, a method similar to \cite{taylor16} is adopted wherein the Richardson number of the flow is adjusted via $g$ using a mass-spring damper control system with a target value for the total TKE $(1/2)\|\boldsymbol{u}\|^2$. With this method, the Richardson and Froude numbers are emergent in the flow, rather than imposed. In the absence of stratification, the TKE in a homogeneous turbulent shear flow grows with time (while the flows scales remain smaller than the domain size). With stratification, mechanisms such as spontaneous shear instabilities that form in the flow due to the layering provide for a downscale TKE flux. A statistically stationary, homogeneous, sheared, stably stratified flow therefore exists at a special equilibrium point where mechanisms generating a downscale TKE flux prevent the growth of TKE that would occur in the flow in the absence of stratification. Put another way, coupling of the fluctuating momentum to the density fields provides for additional energy dissipation mechanisms, which allows the sheared flow to attain a stead-state.

The flow is well-resolved, with isotropic grid spacing and maximum wavenumber $k_{max}$ satisfying $k_{max}\eta\approx 2$. In order to allow for the development of large anisotropic flow scales, a large domain of size $\mathcal{L}_x/\mathcal{L}_y=2$, $\mathcal{L}_x/\mathcal{L}_z=4$, $\mathcal{L}_x/L=40$ was used, where $\mathcal{L}_x,\mathcal{L}_y,\mathcal{L}_z$ are the dimensions of the domain in the $x,y,z,$ directions. In the $x$ direction 1792 grid points were used. Table \ref{table:DNS} summarizes the parameters in the DNS.

\section{Results and Discussion}\label{RandD}
\subsection{Mean-field behavior}
{\vspace{0mm}\begin{figure}
		\centering
		\subfloat[]
		{\begin{overpic}
				[trim = 0mm 60mm 0mm 70mm,scale=0.32,clip,tics=20]{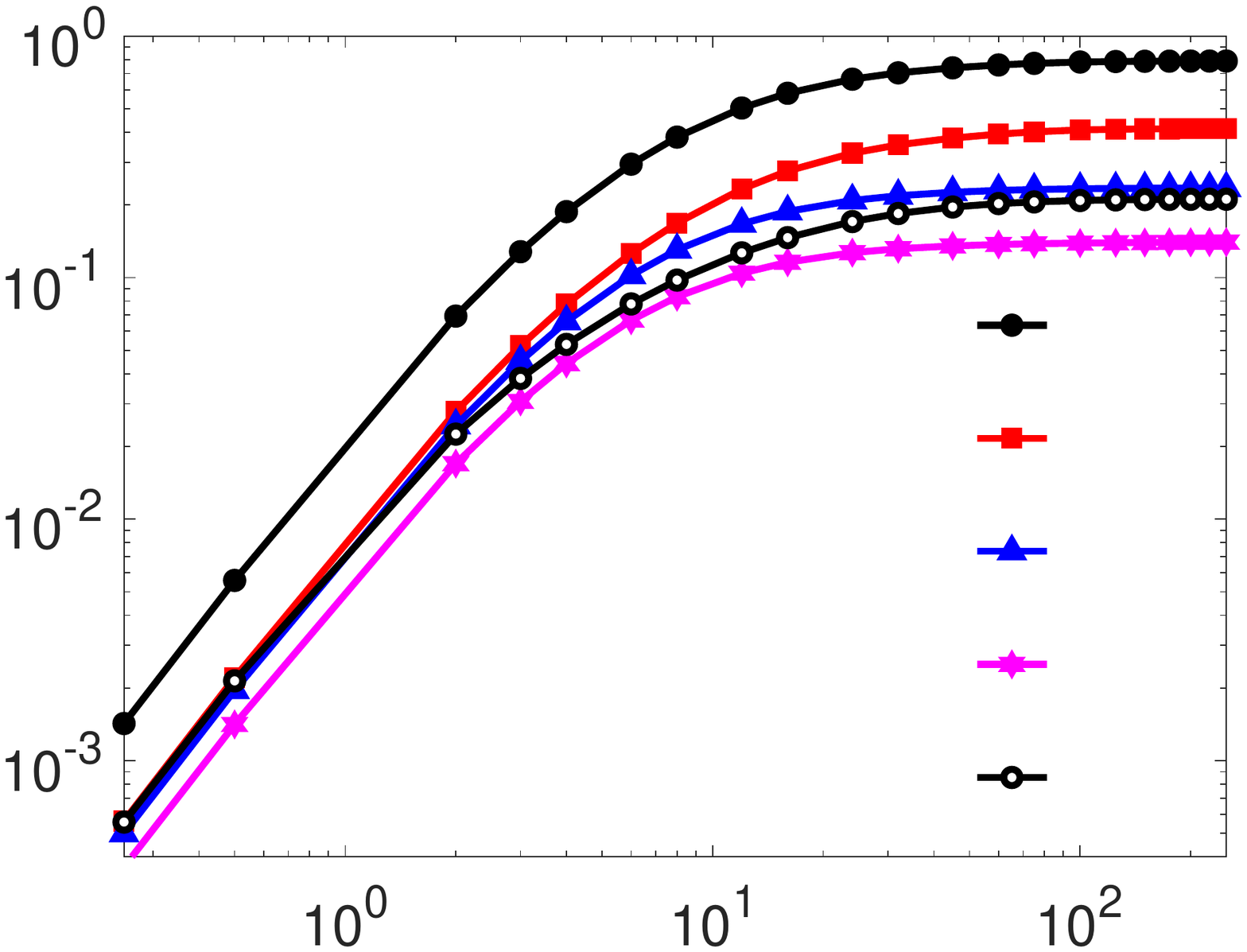}
				\put(100,-2){$\ell/\eta$}
				\put(98,90){\footnotesize$\langle e_K\rangle/\langle E_T\rangle $}
				\put(85,73){\footnotesize$\langle \tau_{xx}\rangle/[2\langle E_T\rangle ]$}
				\put(85,58){\footnotesize$\langle \tau_{yy}\rangle/[2\langle E_T\rangle ]$}				
				\put(85,43){\footnotesize$\langle \tau_{zz}\rangle/[2\langle E_T\rangle ]$}				
				\put(98,27){\footnotesize$\langle e_P\rangle/\langle E_T\rangle$}
		\end{overpic}}
				\subfloat[]
		{\begin{overpic}
				[trim = 0mm 60mm 0mm 70mm,scale=0.32,clip,tics=20]{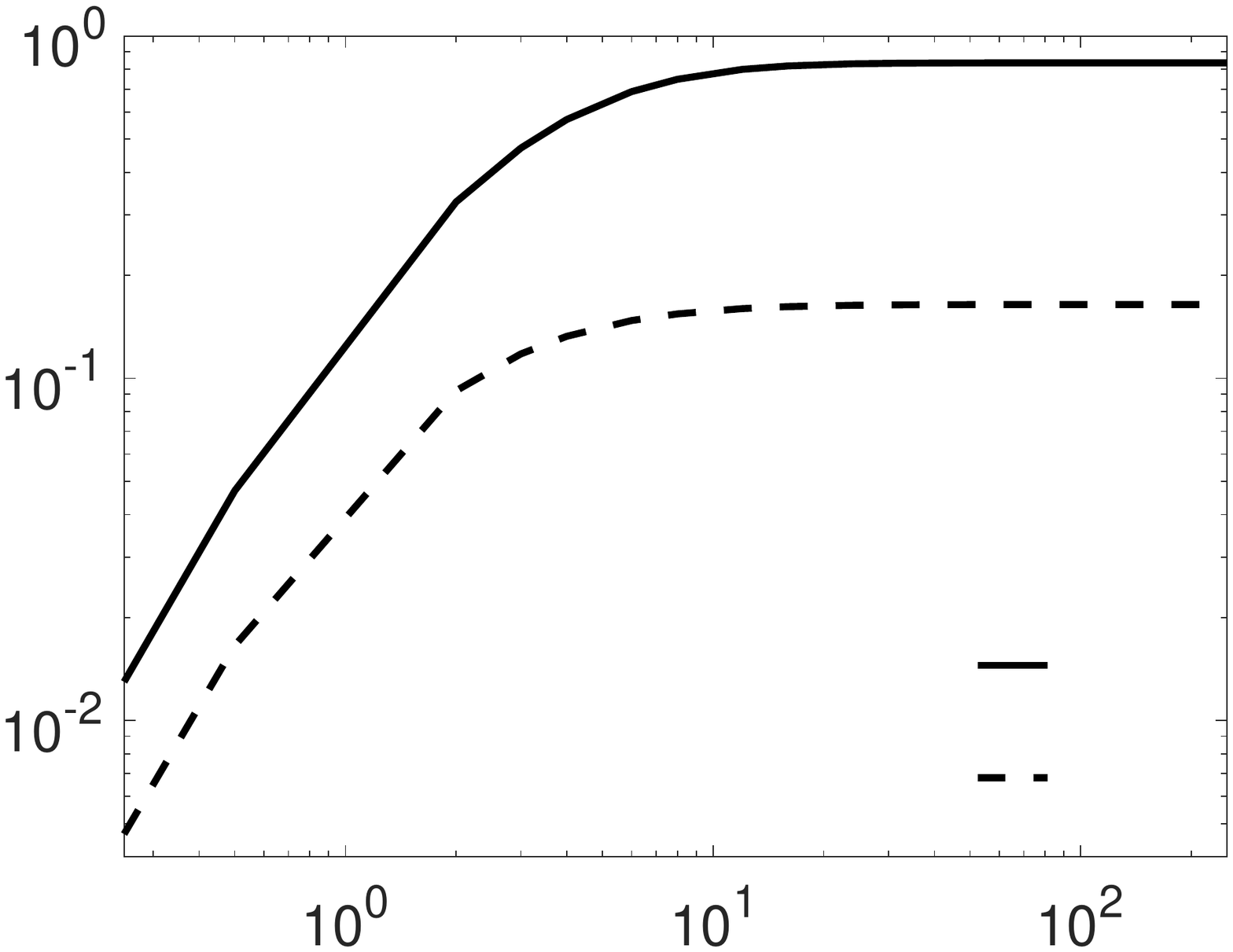}
				\put(100,-2){$\ell/\eta$}
				\put(98,42){\footnotesize$\langle \varepsilon_K\rangle/\langle \epsilon_T\rangle $}
				\put(98,27){\footnotesize$\langle \varepsilon_P\rangle/\langle \epsilon_T\rangle $}
		\end{overpic}}
		\caption{(a) Plot of mean small-scale TKE $\langle e_K\rangle$, TPE $\langle e_P\rangle$, and diagonal components of $\langle\boldsymbol{\tau}\rangle/2$, normalized by total energy $\langle E_T\rangle \equiv\lim_{\ell/\eta\to\infty}[\langle e_K\rangle+\langle e_P\rangle]$, as a function of filter scale $\ell$. (b) Plot of mean small-scale TKE $\langle \varepsilon_K\rangle$ and TPE $\langle \varepsilon_P\rangle$ dissipation rates, normalized by the total turbulent energy dissipation rate $\langle \epsilon_T\rangle\equiv\lim_{\ell/\eta\to\infty}[\langle \varepsilon_K\rangle+\langle \varepsilon_P\rangle]$.
		} 
		\label{TKE_PE_plot}
\end{figure}}
We begin by considering the behavior of $\langle e_K\rangle$, $\langle e_P\rangle$ and the diagonal components of $\langle\boldsymbol{\tau}\rangle/2$ (which correspond to the TKE associated with different components of $\boldsymbol{u}$) as a function of filter length $\ell$ in order to understand how energy is partitioned in the flow at different scales. The results in figure \ref{TKE_PE_plot}(a) show that at all scales, $\langle e_K\rangle>\langle e_P\rangle$, with $\langle e_K\rangle/\langle e_P\rangle=O(10)$ at the large scales of the flow. The flow is therefore far from a state of equipartition of large-scale energy among the TKE and TPE fields, unlike the behavior that is thought to emerge for strongly stratified flows with $Fr\ll 1$ \citep{billant01}. 
Below the Corrsin scale $\ell_C=?$, the difference between $\langle e_K\rangle$ and $\langle e_P\rangle$ reduces, but remains significant even for $\ell/\eta\ll 1$. When considering the components $\langle\boldsymbol{\tau}\rangle/2$, it is seen that although the total TKE is much larger than the TPE, the TKE associated with particular components of the velocity field is comparable to the TPE. In particular, $\langle \tau_{yy}\rangle/2\approx \langle e_P\rangle$ and $\langle \tau_{zz}\rangle/2\approx (2/3)\langle e_P\rangle$ at the large scales. As such, even though the energy contained in the TPE field is small compared to that in the total TKE field, it is of the same order as that contained in the spanwise ($y$) and vertical ($z$) directions of the flow, and therefore plays an important energetic role in the system. In terms of scaling, $\langle e_K\rangle$ and $\langle e_P\rangle$ both show the expected behavior $\langle e_K\rangle\propto \ell^2$ and $\langle e_P\rangle\propto \ell^2$ at $\ell/\eta\leq O(1)$ where viscous effects lead to smooth velocity and density fields. Outside of this a well defined scaling regime does not emerge, due to the limited Reynolds number of the flow (as well as the fact that $\ell_C/\eta$ is too small for an inertially dominated regime to emerge).

In figure \ref{TKE_PE_plot}(b) we consider the mean small-scale turbulent kinetic $\langle \varepsilon_K\rangle$ and potential $\langle \varepsilon_P\rangle$ energy dissipation rates. For $\ell/\eta\to\infty$ these satisfy $\langle \varepsilon_K\rangle\to \langle \epsilon_K\rangle$ and $\langle \varepsilon_P\rangle\to \langle \epsilon_P\rangle$, and the results show that these are in the ratio $\langle \epsilon_K\rangle/\langle \epsilon_P\rangle\approx 5$. Both $\langle \varepsilon_K\rangle$ and $\langle \varepsilon_P\rangle$ are only approximately independent of $\ell$ down to $\ell/\eta= O(10)$, consistent with the usual observation that the Kolmogorov scale underestimates the scale at which viscous forces become important \citep{pope}.

{\vspace{0mm}\begin{figure}
		\centering
		{\begin{overpic}
				[trim = 0mm 60mm 0mm 70mm,scale=0.32,clip,tics=20]{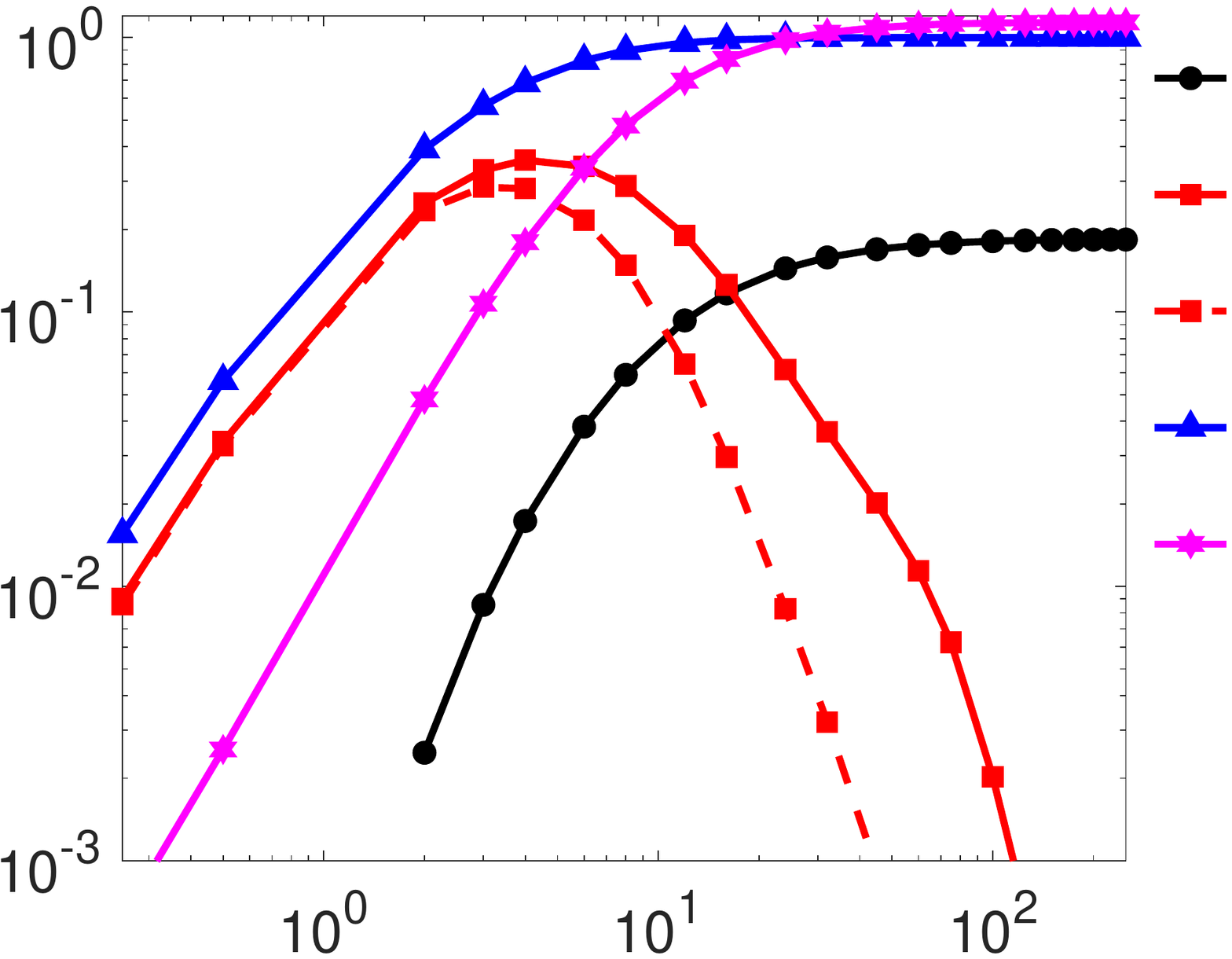}
				\put(80,-3){$\ell/\eta$}
				\put(167,121){$-\langle \mathcal{B}\rangle/\langle\epsilon_K\rangle$}
				\put(167,105){$\langle \Pi_K\rangle/\langle\epsilon_K\rangle$}
				\put(167,89){$\langle \Pi_K^F\rangle/\langle\epsilon_K\rangle$}
				\put(167,74){$-\langle \varepsilon_K\rangle/\langle\epsilon_K\rangle$}
				\put(167,59){$\langle \mathcal{F}_K\rangle/\langle\epsilon_K\rangle$}
		\end{overpic}}
		\caption{Plot of terms in the average small-scale TKE budget equation \eqref{TKE_budget_2}.} 
		\label{TKE_budgets}
\end{figure}}

We now turn to consider the contributions to the equations governing $\langle e_K\rangle$ and $\langle e_P\rangle$. The results in figure \ref{TKE_budgets} show that the flow exhibits the $\ell/\eta\to\infty$ asymptotic behavior $\langle\mathcal{F}_K\rangle^\infty\sim-\langle\mathcal{B}\rangle^\infty+\langle\epsilon_K\rangle$ for $\ell/\eta \gtrsim O(100)$. As $\ell/\eta$ is decreased below $O(100)$, $\langle \Pi_K\rangle$ begins to increase and becomes positive, indicating a downscale flux of TKE, as has previously been observed for stratified turbulent flows \citep{riley03,waite06,lindborg05,lindborg06,brethouwer07}. However, it does not become significant until $\ell/\eta \lesssim O(10)$. Moreover, when it reaches its peak value, it does not exhibit the behavior $\langle \Pi_K\rangle/\langle\epsilon_K\rangle\sim 1$ which would be expected for non-stratified isotropic turbulence. This is because $\langle \Pi_K\rangle$ cannot grow significantly until $\ell$ is small enough such that $\langle\mathcal{F}_K\rangle\ll\langle\mathcal{F}_K\rangle^\infty$, which only occurs for $\ell\leq O(\ell_C)$. Once this regime is obtained, then since $\ell_C<\ell_O$ the TKE balance becomes $\langle\Pi_K\rangle\sim\langle\varepsilon_K\rangle$, a behavior that can be observed in figure \ref{TKE_budgets} to be approached at small scales. However, in the present flow, this behavior does not give rise to an inertial TKE cascade since at the scales where $\langle\Pi_K\rangle\sim\langle\varepsilon_K\rangle$, $\langle\varepsilon_K\rangle$ is a decreasing function of $\ell$. In order to observe an inertial TKE cascade with $\langle\Pi_K\rangle\sim\langle\epsilon_K\rangle$ would require considering a flow possessing a range of scales $\ell_O>\ell_C\gg\ell\gg\eta$. The results in figure \ref{TKE_budgets} also show that the contribution to the TKE flux coming from the filtered field dynamics, i.e. $\langle\Pi_K^F\rangle$, makes a significant contribution to $\langle\Pi_K\rangle$ at scales where $\langle\Pi_K\rangle$ plays a significant role in the small-scale TKE budget equation, but also that the contribution involving the sub-grid fields $\langle\Pi_K^{SG}\rangle$ is also significant, just as for isotropic turbulence \citep{johnson20,johnson21}. At sufficiently small $\ell$, $\langle\Pi_K\rangle\approx \langle\Pi_K^F\rangle$, consistent with the exact limiting behavior $\lim_{\ell/\eta\to 0}\Pi_K^F\to\Pi_K$.

{\vspace{0mm}\begin{figure}
		\centering
		{\begin{overpic}
				[trim = 0mm 60mm 0mm 70mm,scale=0.32,clip,tics=20]{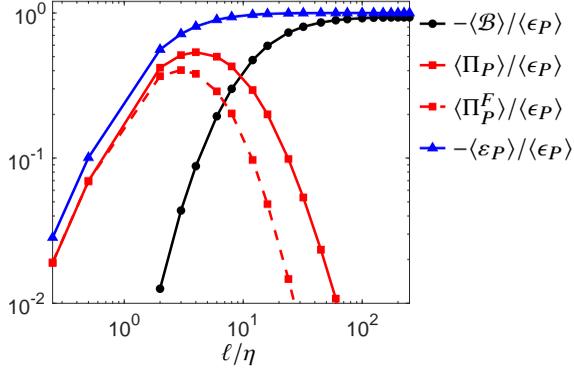}   
				\put(80,-3){$\ell/\eta$}
				\put(167,121){$-\langle \mathcal{B}\rangle/\langle\epsilon_P\rangle$}
				\put(167,105){$\langle \Pi_P\rangle/\langle\epsilon_P\rangle$}
				\put(167,89){$\langle \Pi_P^F\rangle/\langle\epsilon_P\rangle$}
				\put(167,74){$-\langle \varepsilon_P\rangle/\langle\epsilon_P\rangle$}
		\end{overpic}}	
		\caption{Plot of terms in the average small-scale TPE budget equation \eqref{PE_budget_2}.} 
		\label{PE_budgets}
\end{figure}}
\FloatBarrier

Concerning the TPE behavior, the results in figure \ref{PE_budgets} exhibit the $\ell/\eta\to \infty$ asymptotic behavior $\langle\mathcal{B}\rangle^\infty\sim-\langle\epsilon_P\rangle$ for $\ell/\eta \gtrsim O(100)$. The behavior of $\langle \Pi_P\rangle$ is very similar to that of $\langle \Pi_K\rangle$ in that $\langle \Pi_P\rangle$ does not become significant until $\ell/\eta \lesssim O(10)$, and does not exhibit the behavior $\langle \Pi_P\rangle/\langle\epsilon_P\rangle\sim 1$ that would be expected for a passive scalar advected in isotropic turbulence. Whereas the growth of $\langle \Pi_K\rangle$ is constrained in SSST to scales $\ell\leq O(\ell_C)$, as discussed previously, the growth of $\langle \Pi_P\rangle$ is constrained in SSST to scales $\ell\leq O(\ell_O)$. In our DNS these two ranges are similar, and so the range of scales over which $\langle \Pi_K\rangle$ and $\langle \Pi_P\rangle$ are active is similar. However, in an SSST flow with $Ri\lll1$ and $Re\ggg 1$ this could lead to an interesting scenario where $\langle \Pi_P\rangle/\langle\epsilon_P\rangle\sim 1$ while $\langle \Pi_K\rangle/\langle\epsilon_K\rangle\approx 0$ for $\ell_O\gg\ell\gg\ell_C$. The results in figure \ref{PE_budgets} also show that like the TKE results, the contribution to the TPE flux coming from the filtered field dynamics, i.e. $\langle\Pi_P^F\rangle$, makes a significant contribution to $\langle\Pi_P\rangle$ at scales where $\langle\Pi_P\rangle$ plays a significant role in the small-scale TPE budget equation, but that the contribution involving the sub-grid fields $\langle\Pi_P^{SG}\rangle$ is also significant. At sufficiently small $\ell$, $\langle\Pi_P\rangle\approx \langle\Pi_P^F\rangle$, consistent with the exact limiting behavior $\lim_{\ell/\eta\to 0}\Pi_P^F\to\Pi_P$.

\subsection{Fluctuations about the mean-field}
{\vspace{0mm}\begin{figure}
		\centering
		\subfloat[]
		{\begin{overpic}
				[trim = 0mm 60mm 0mm 75mm,scale=0.32,clip,tics=20]{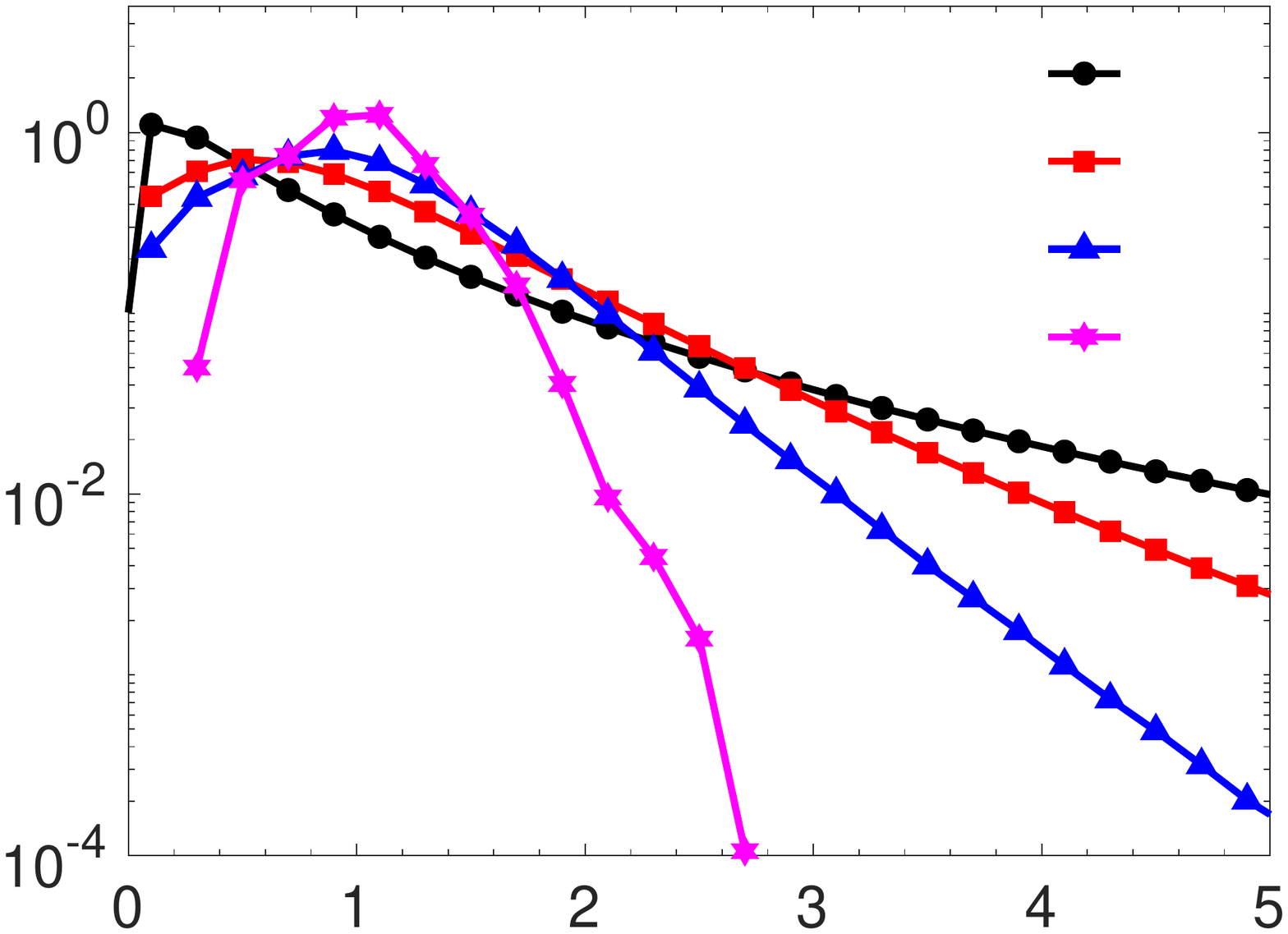}
				\put(83,0){$e_K/\langle e_K\rangle$}
				\put(-3,60){\rotatebox{90}{PDF}}
				\put(108,120){$\ell/\eta=0.25$}
				\put(108,109){$\ell/\eta=6$}		
				\put(108,97){$\ell/\eta=16$}
				\put(108,85){$\ell/\eta=60$}
		\end{overpic}}
				\subfloat[]
		{\begin{overpic}
				[trim = 0mm 60mm 0mm 75mm,scale=0.32,clip,tics=20]{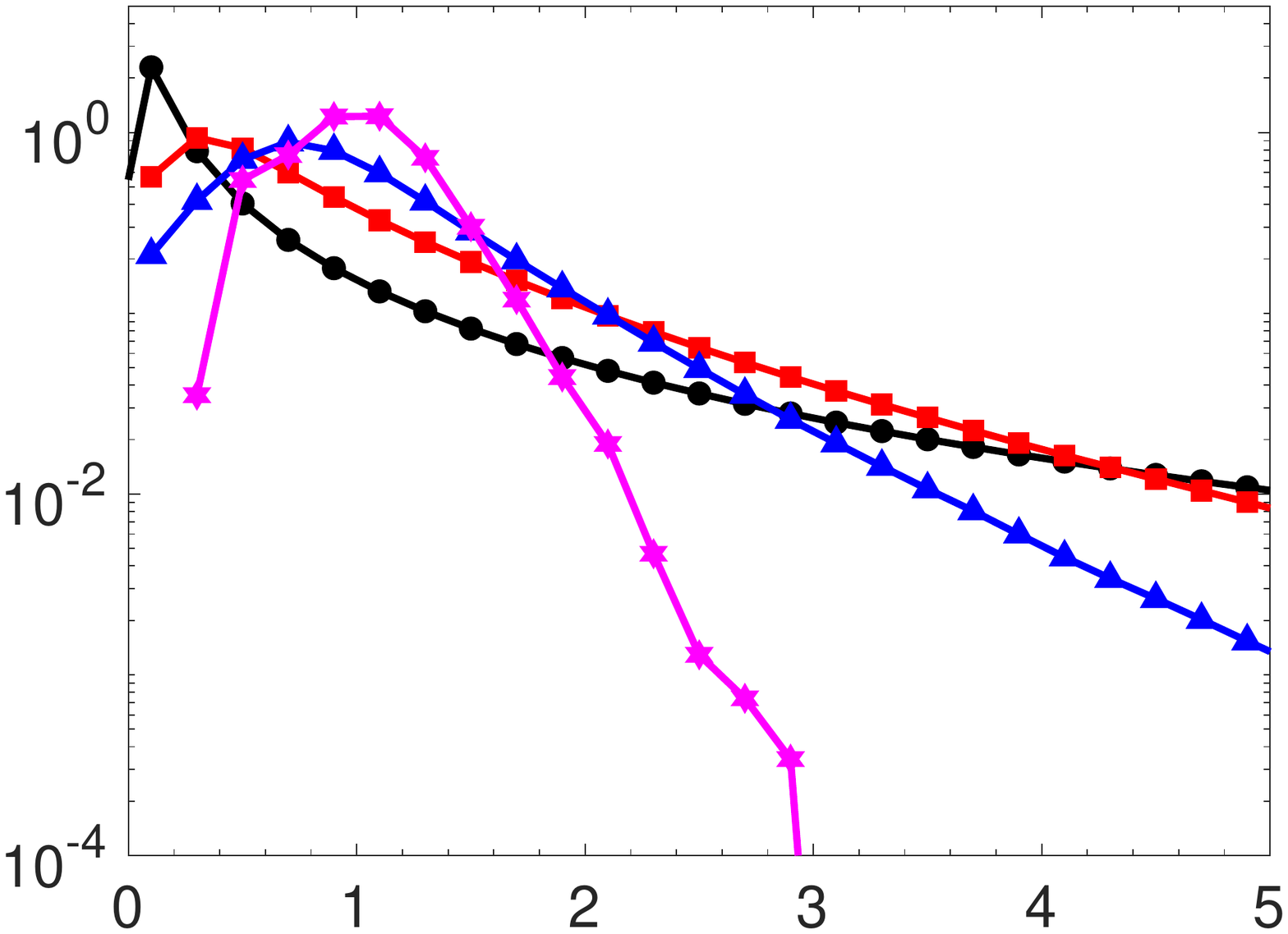}
				\put(83,0){$e_P/\langle e_P\rangle$}
				\put(-3,60){\rotatebox{90}{PDF}}
		\end{overpic}}\\
						\subfloat[]
		{\begin{overpic}
						[trim = 0mm 60mm 0mm 75mm,scale=0.32,clip,tics=20]{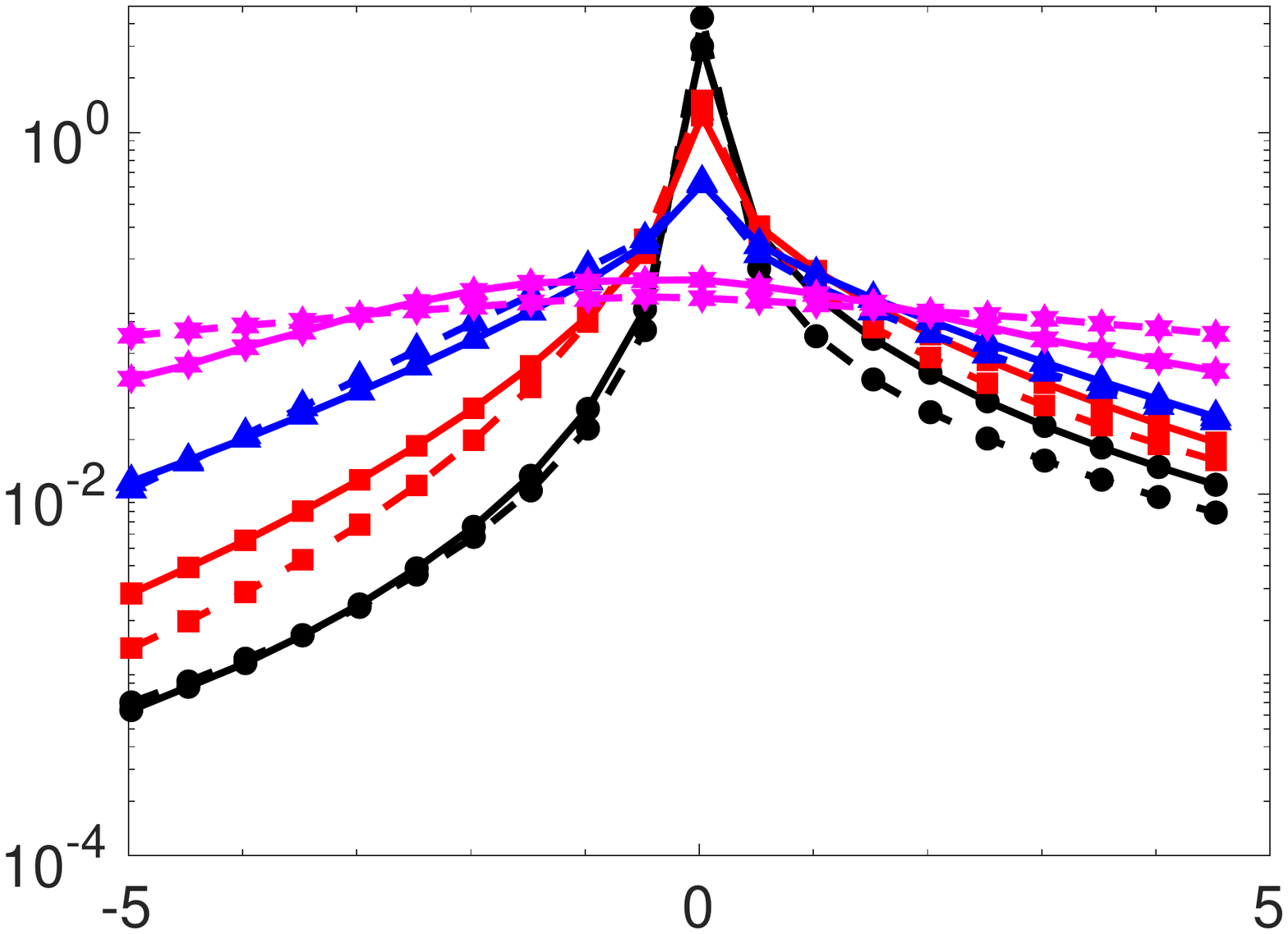}
				\put(99,0){$\xi$}
				\put(-3,60){\rotatebox{90}{PDF}}
		\end{overpic}}
						\subfloat[]		
				{\begin{overpic}
				[trim = 0mm 60mm 0mm 75mm,scale=0.32,clip,tics=20]{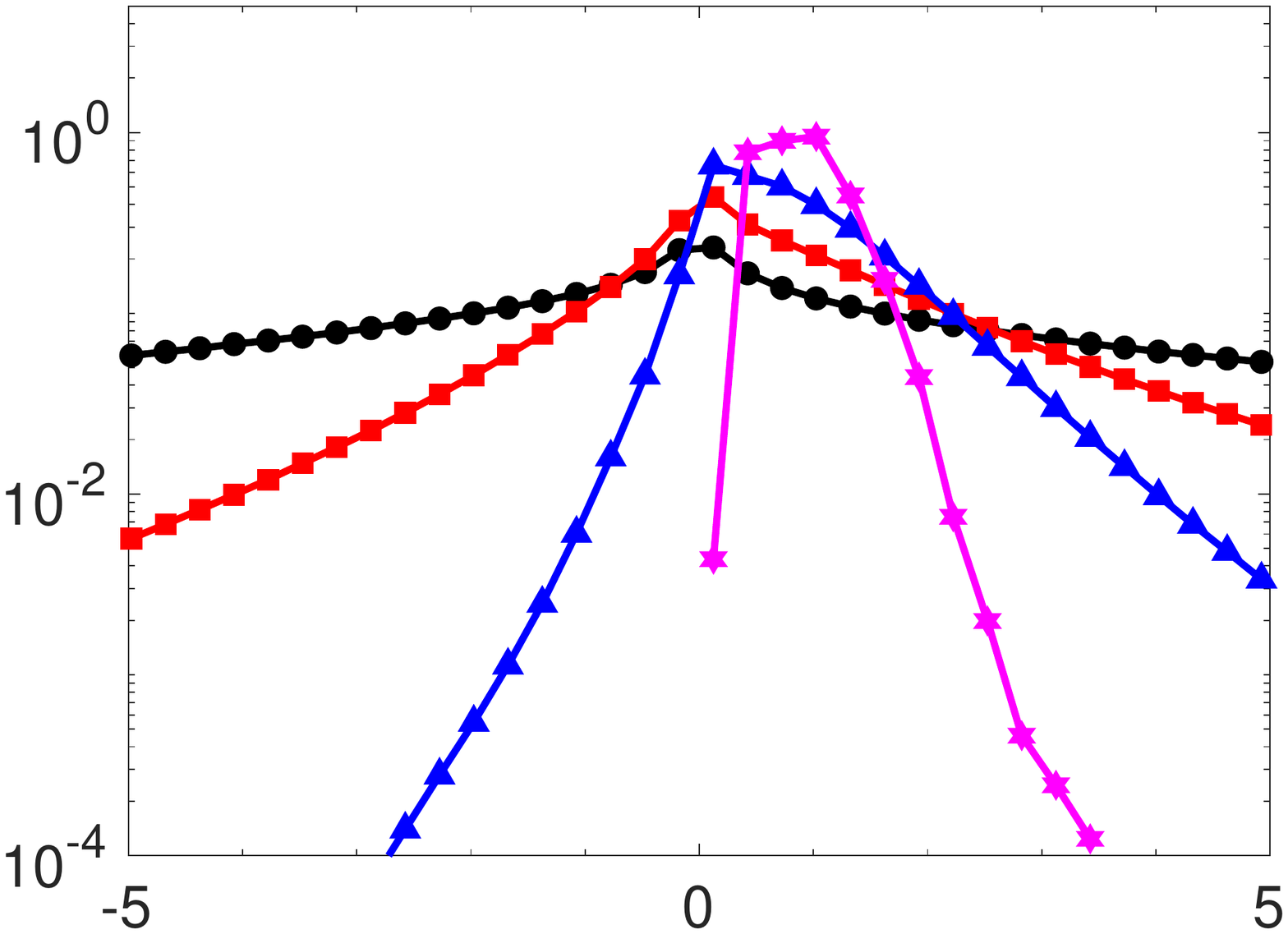}
				\put(89,0){$\mathcal{B}/\langle\mathcal{B}\rangle$}
				\put(-3,60){\rotatebox{90}{PDF}}				
		\end{overpic}}
		\caption{Plots of PDF of (a) $e_K/\langle e_K\rangle$, (b) $e_P/\langle e_P\rangle$ and (d) $\mathcal{B}/\langle\mathcal{B}\rangle$ for different filter lengths $\ell$. In plot (c), the solid lines correspond to the PDFs of $\Pi_K/\langle\Pi_K\rangle$ and the dashed lines correspond to the PDFs of $\Pi_P/\langle\Pi_P\rangle$. Different colors/symbols correspond to different $\ell/\eta$ as indicated by the legend in plot (a).} 
		\label{PDF_various}
\end{figure}}
\FloatBarrier

In order to understand the energetics of the flow beyond its mean-field behavior, we will now consider the Probability Density Functions (PDFs) of various quantities. We begin in figure \ref{PDF_various}(a),(b) with the PDFs of the normalized small-scale energies $e_K/\langle e_K\rangle$ and $e_P/\langle e_P\rangle$ for different filtering lengths $\ell/\eta$. At the small scales, the PDFs are highly non-Gaussian, such that the TKE and TPE fields exhibit frequent large fluctuations about their mean-field behaviour. As the filter length increases, the PDFs approach a Gaussian shape, associated with the dominance of the linear mean-shear and buoyancy forces at the large scales of the flow. The results also show that the TPE field exhibits stronger fluctuations from the mean-field behavior than the TKE field. This is consistent with the fact that scalar fields are known to be more intermittent in turbulent flows because they lack the pressure gradient in their dynamics that regulates large fluctuations in the velocity gradient field.

In figure \ref{PDF_various}(c) we show the PDFs of $\Pi_K/\langle\Pi_K\rangle$ and $\Pi_P/\langle\Pi_P\rangle$ for different filtering lengths $\ell/\eta$. At larger scales, the PDFs have a large variance, showing that at these scales, the fluxes of TKE and TPE can significantly exceed their mean values. The PDFs are also close to being symmetric, so that the probability of upscale or downscale fluxes TKE are similar, and the same for the TPE. As one moves to smaller scales where $\langle\Pi_K\rangle$ and $\langle\Pi_P\rangle$ both increase, the probability of large fluctuations about the mean-field behavior reduce. However, there is still a significant probability of observing regions of the flow where the local values of $\Pi_K$ and $\Pi_P$ significantly exceed their mean values in magnitude. As $\ell$ is reduced, the mode of the PDF remains close to zero, while the mean and skewness both become positive, with the probability of downscale fluxes significantly exceeding that of upscale fluxes. The PDFs for TKE and TPE are similar in shape, and at scales where $\langle\Pi_K\rangle$ and $\langle\Pi_P\rangle$ play an important role in the TKE and TPE scale-wise budgets, the fluctuations of $\Pi_K$ about $\langle\Pi_K\rangle$ are larger than those of $\Pi_P$ about $\langle\Pi_P\rangle$.
{\vspace{0mm}\begin{figure}
		\centering
		\subfloat[]
		{\begin{overpic}
				[trim = 0mm 50mm 0mm 70mm,scale=0.32,clip,tics=20]{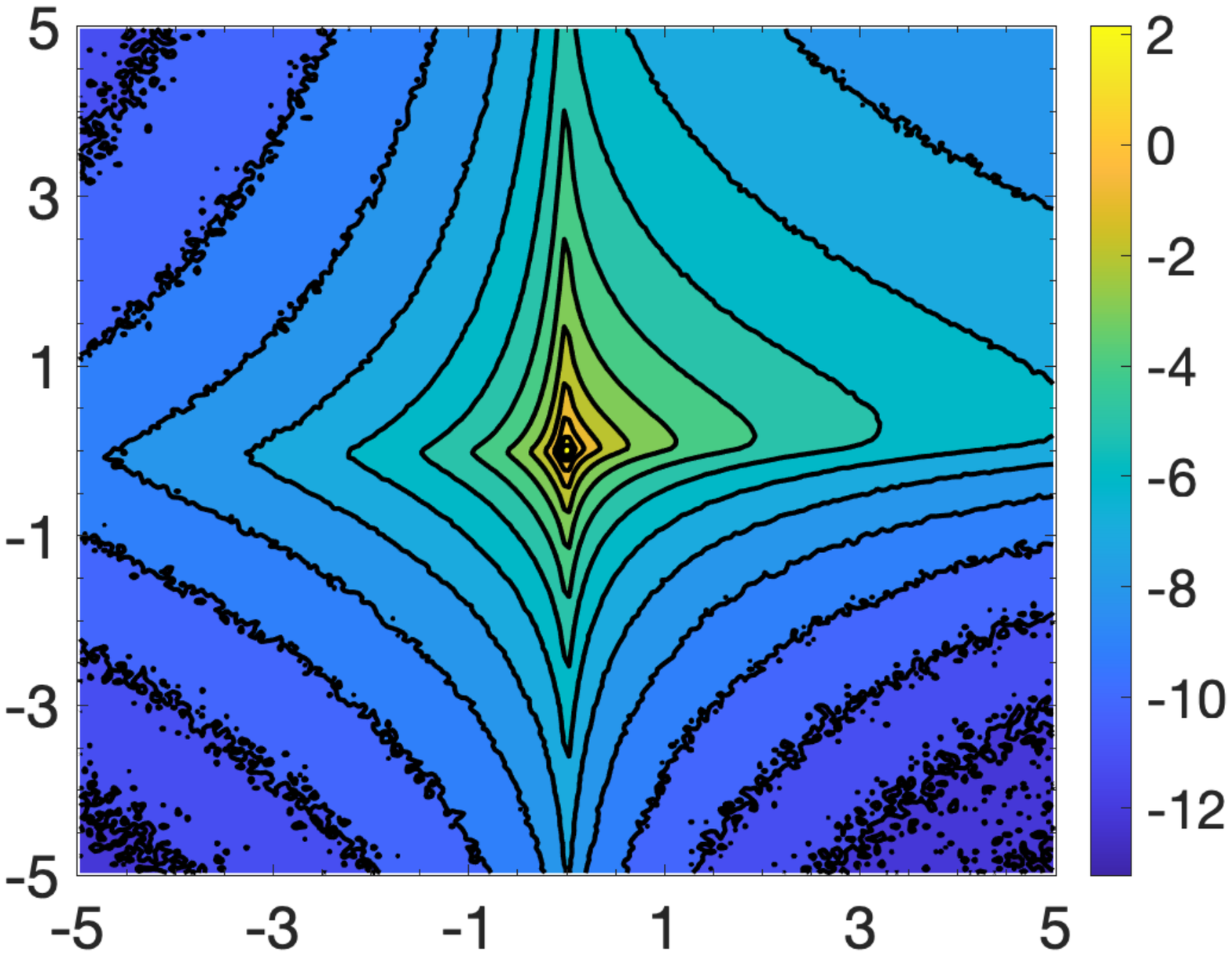}
				\put(80,8){$\Pi_P/\langle\Pi_P\rangle$}
				\put(5,65){\rotatebox{90}{$\Pi_K/\langle\Pi_K\rangle$}}
		\end{overpic}}
				\subfloat[]
		{\begin{overpic}
				[trim = 0mm 50mm 0mm 70mm,scale=0.32,clip,tics=20]{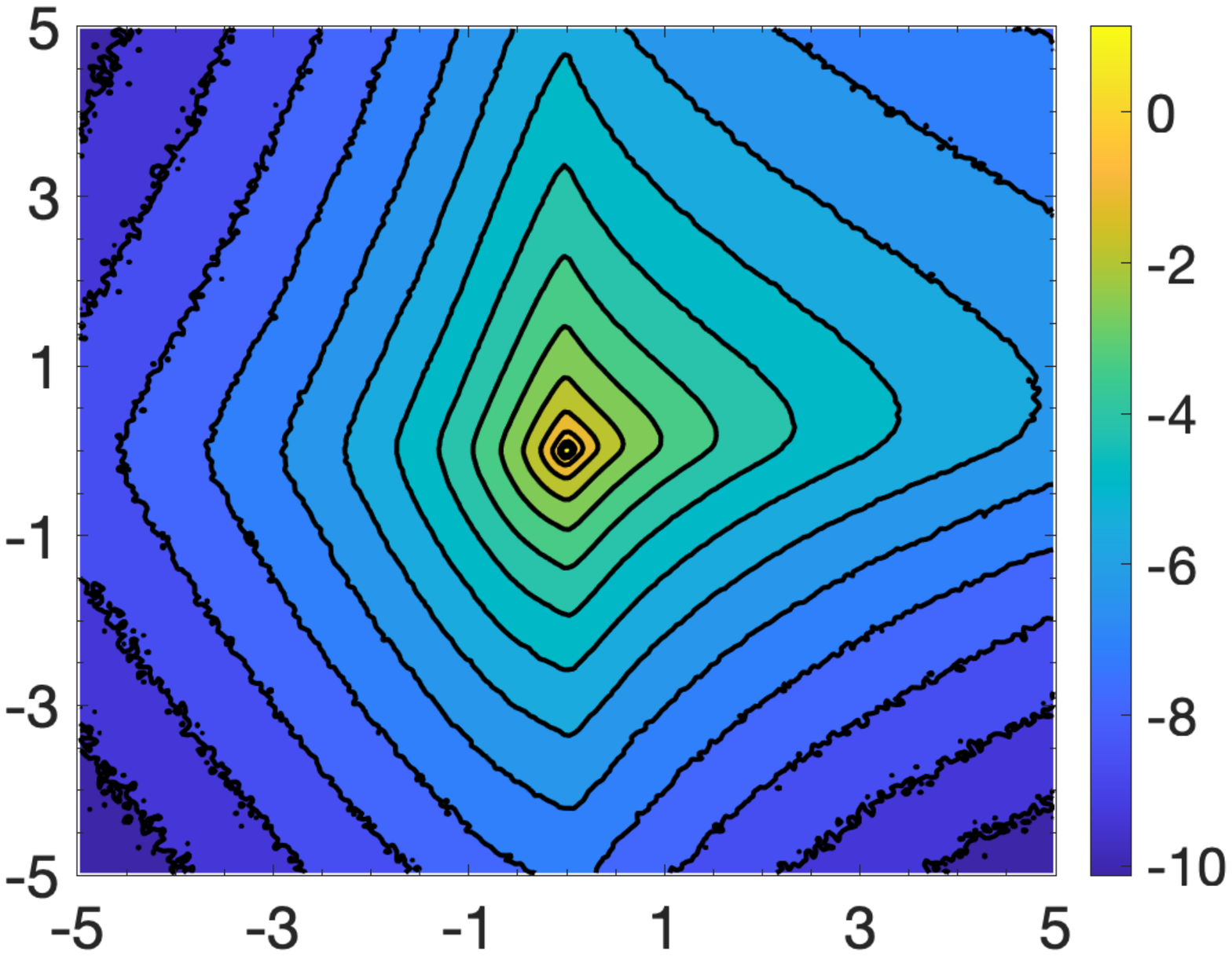}
				\put(80,8){$\Pi_P/\langle\Pi_P\rangle$}
				\put(5,65){\rotatebox{90}{$\Pi_K/\langle\Pi_K\rangle$}}			
		\end{overpic}}\\
			\subfloat[]
				{\begin{overpic}
				[trim = 0mm 50mm 0mm 70mm,scale=0.32,clip,tics=20]{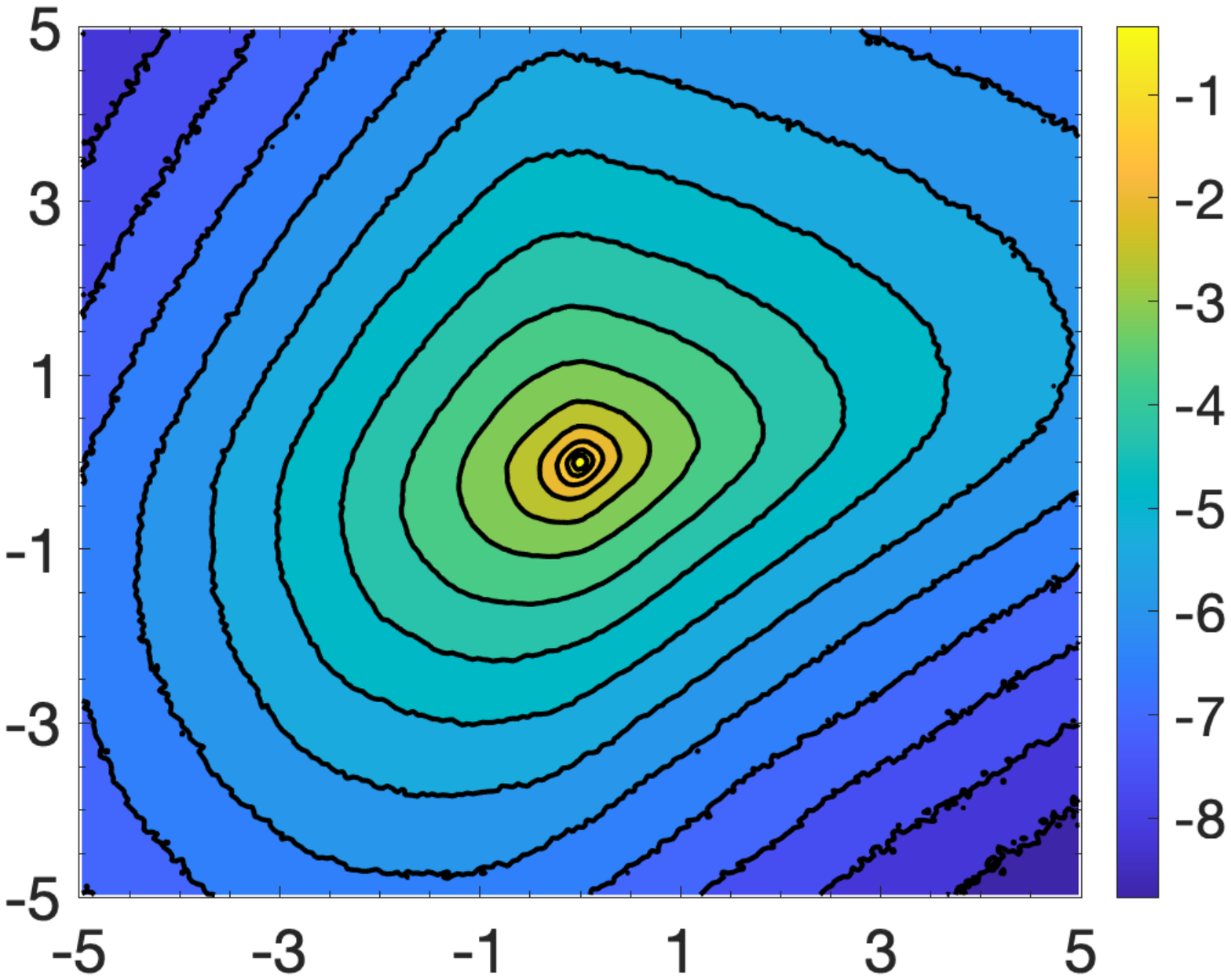}
				\put(80,8){$\Pi_P/\langle\Pi_P\rangle$}
				\put(5,65){\rotatebox{90}{$\Pi_K/\langle\Pi_K\rangle$}}
		\end{overpic}}
				\subfloat[]
		{\begin{overpic}
				[trim = 0mm 50mm 0mm 70mm,scale=0.32,clip,tics=20]{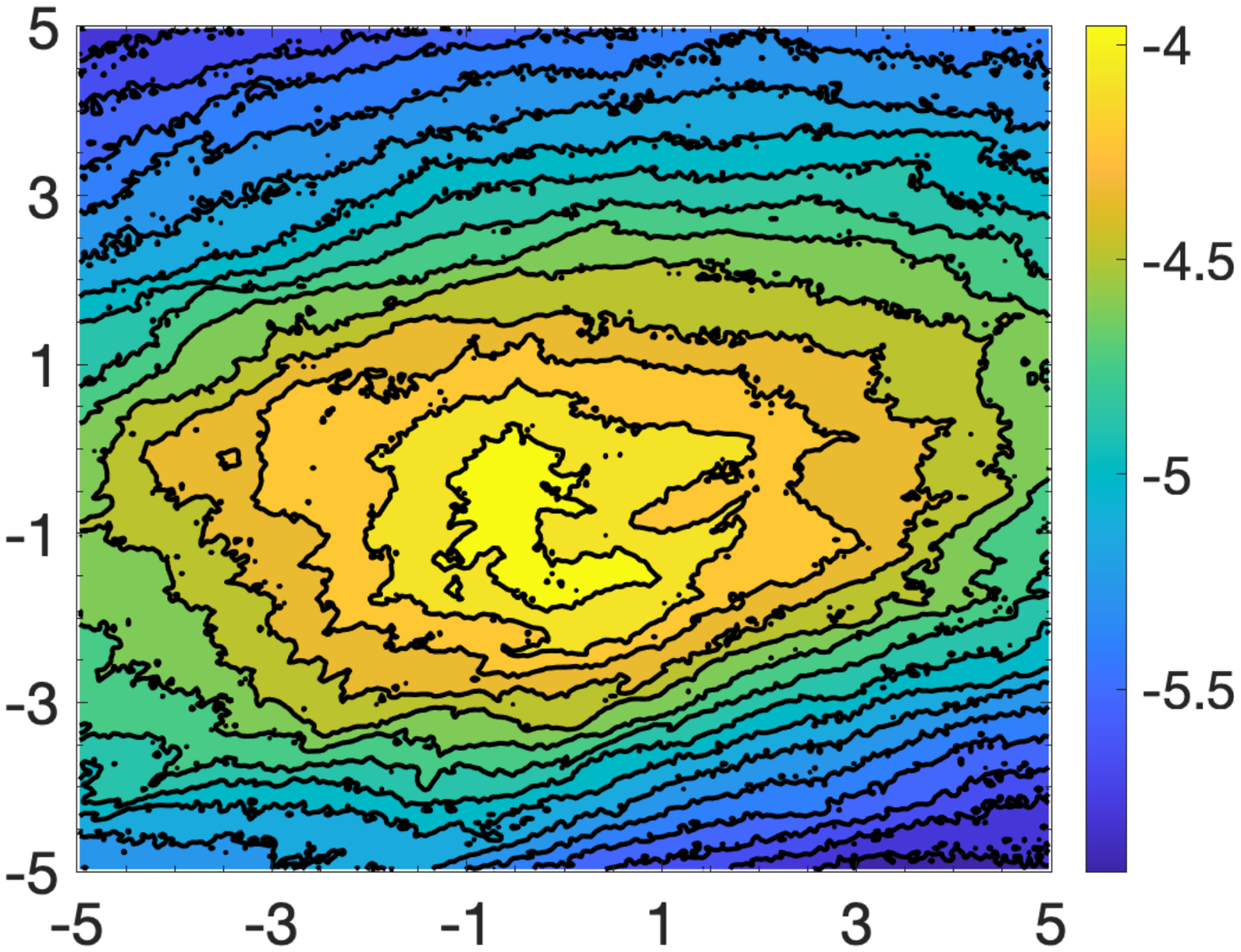}
				\put(80,8){$\Pi_P/\langle\Pi_P\rangle$}
				\put(5,65){\rotatebox{90}{$\Pi_K/\langle\Pi_K\rangle$}}				
		\end{overpic}}
		\caption{Contour plot of the logarithm of the joint PDF of $\Pi_P/\langle\Pi_P\rangle$ and $\Pi_K/\langle\Pi_K\rangle$ for (a) $\ell/\eta=0.25$, (b) $\ell/\eta=6$, (c) $\ell/\eta=16$, (d) $\ell/\eta=60$. Colors correspond to the logarithm of the PDF.} 
		\label{JPF_PiK_PiP}
\end{figure}}
In figure \ref{PDF_various}(d) we show the PDFs of $\mathcal{B}/\langle\mathcal{B}\rangle$ for different filtering lengths $\ell/\eta$. At larger scales where $-\langle\mathcal{B}\rangle$ is the dominant source term in the mean TPE budget, the fluctuations about $\langle\mathcal{B}\rangle$ are not that strong, and remarkably, the probability to observe $\mathcal{B}/\langle\mathcal{B}\rangle<0$, which would correspond to conversion of TPE to TKE, is zero. For smaller scales, the probability to observe $\mathcal{B}/\langle\mathcal{B}\rangle<0$ becomes finite and increases with decreasing $\ell$. The significant fluctuations of $\mathcal{B}$ about its mean value indicate that the effects of buoyancy could be felt at scales considerably smaller than the mean-field description suggests.

In figure \ref{JPF_PiK_PiP} we show the joint PDFs of $\Pi_P/\langle\Pi_P\rangle$ and $\Pi_K/\langle\Pi_K\rangle$ for different filtering lengths $\ell/\eta$. The results reveal that there is a weak positive correlation between $\Pi_P$ and $\Pi_K$ at larger scales, which becomes stronger as $\ell/\eta$ is increased. Nevertheless, even in the range of scales where the TKE and TPE mean fluxes are strongest, the positive correlation between $\Pi_P$ and $\Pi_K$ is not very strong. Indeed, the probability of observing a downscale TKE flux with an upscale TPE flux and vice-versa, is quite significant. This relatively weak correlation can be understood in view of the discussion in \S\ref{FMech} regarding the mechanisms governing the TKE and TPE fluxes. It is common to model TKE and TPE fluxes using similar closures but with a coefficient that accounts for their differences only through the Prandtl number effects. However, the weak correlation between $\Pi_K$ and $\Pi_P$ revealed in figure \ref{JPF_PiK_PiP} shows that such an approach is not suitable, and that the closures need to reflect to some extent the different physical processes dominating these fluxes, which leads to the relatively weak correlation between $\Pi_K$ and $\Pi_P$.
{\vspace{0mm}\begin{figure}
		\centering
		\subfloat[]
		{\begin{overpic}
				[trim = 0mm 50mm 0mm 70mm,scale=0.32,clip,tics=20]{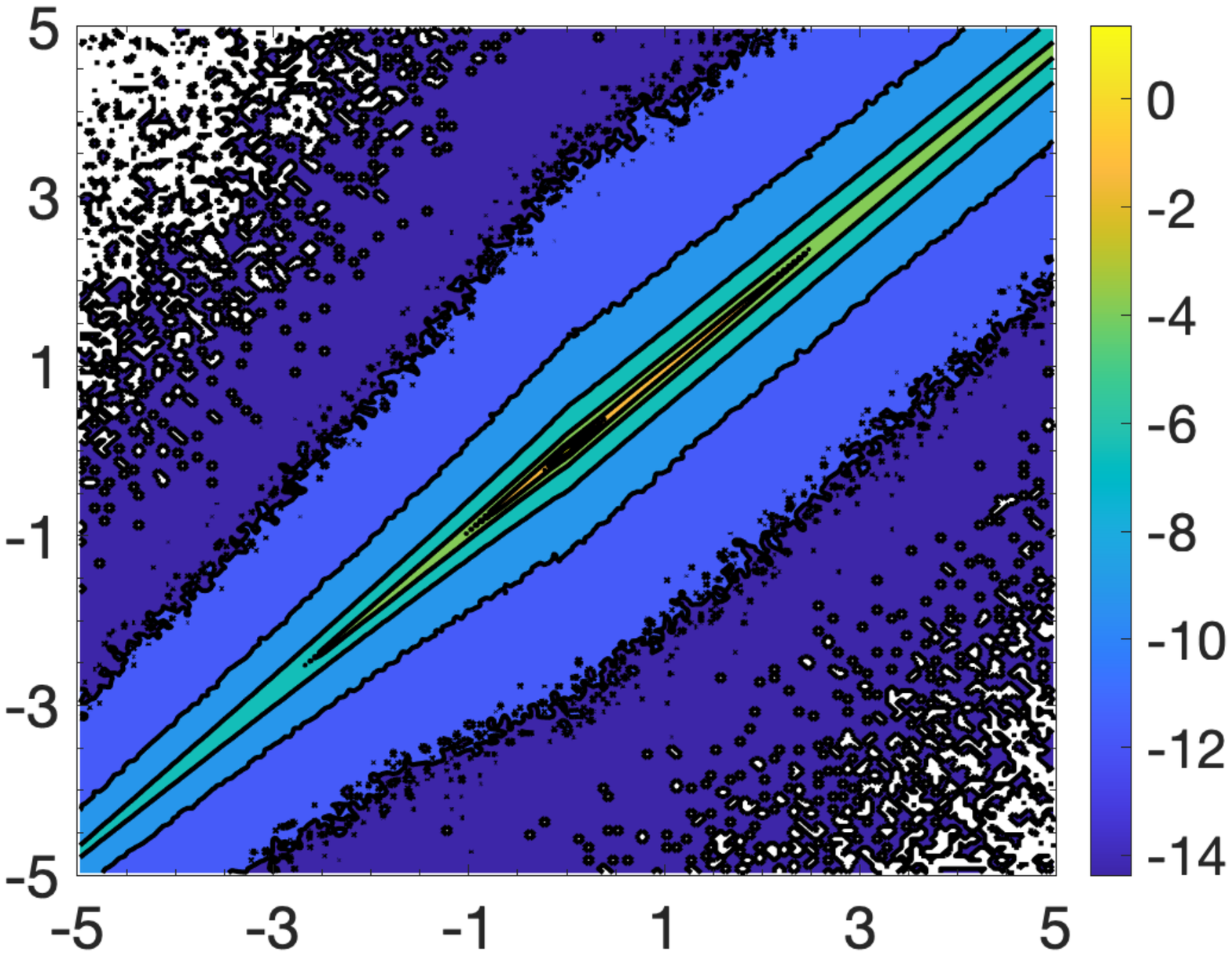}
				\put(80,8){$\Pi_K^F/\langle\Pi_K^F\rangle$}
				\put(5,65){\rotatebox{90}{$\Pi_K/\langle\Pi_K\rangle$}}
		\end{overpic}}
				\subfloat[]
		{\begin{overpic}
				[trim = 0mm 50mm 0mm 70mm,scale=0.32,clip,tics=20]{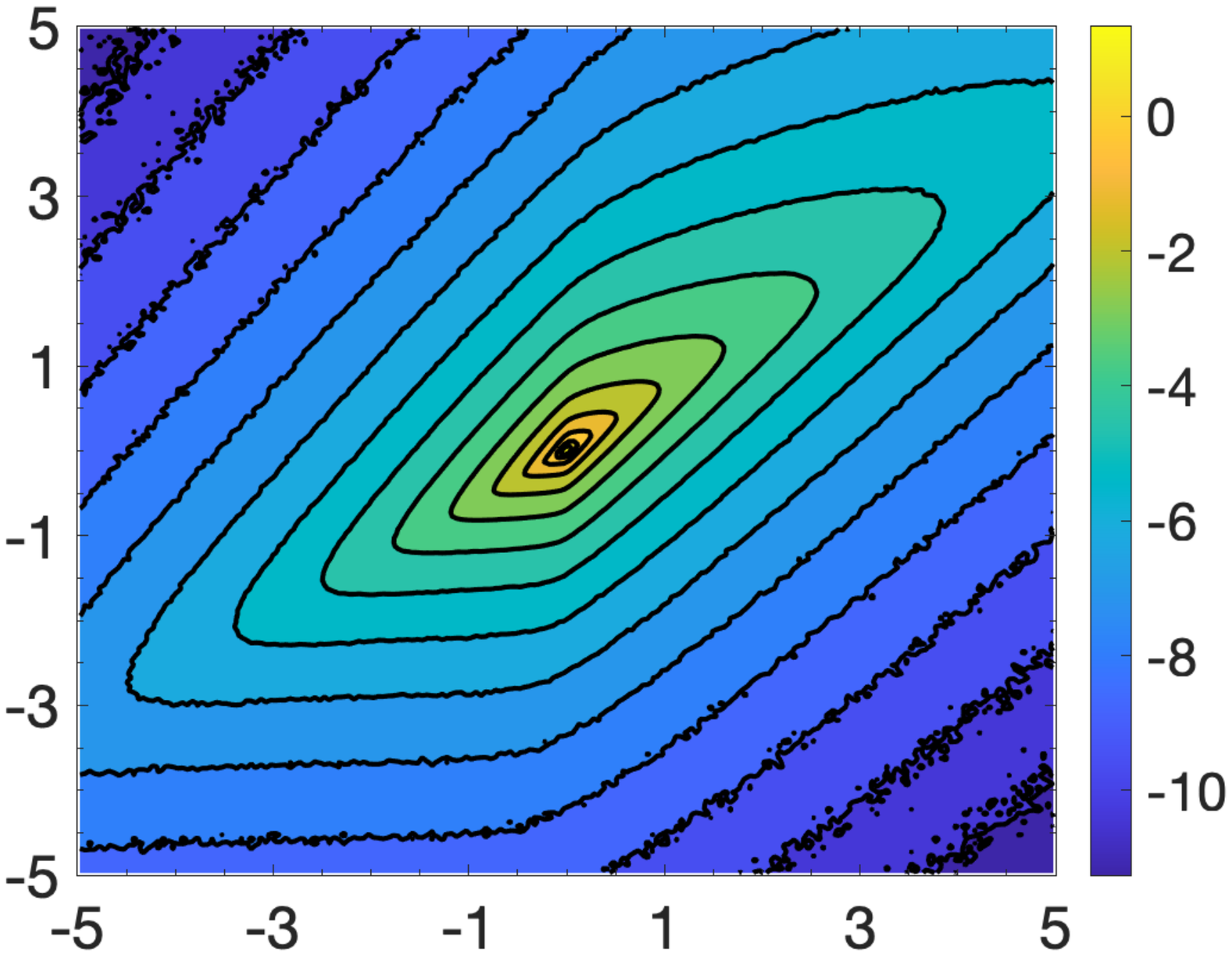}
				\put(80,8){$\Pi_K^F/\langle\Pi_K^F\rangle$}
				\put(5,65){\rotatebox{90}{$\Pi_K/\langle\Pi_K\rangle$}}		
		\end{overpic}}\\
			\subfloat[]
				{\begin{overpic}
				[trim = 0mm 50mm 0mm 70mm,scale=0.32,clip,tics=20]{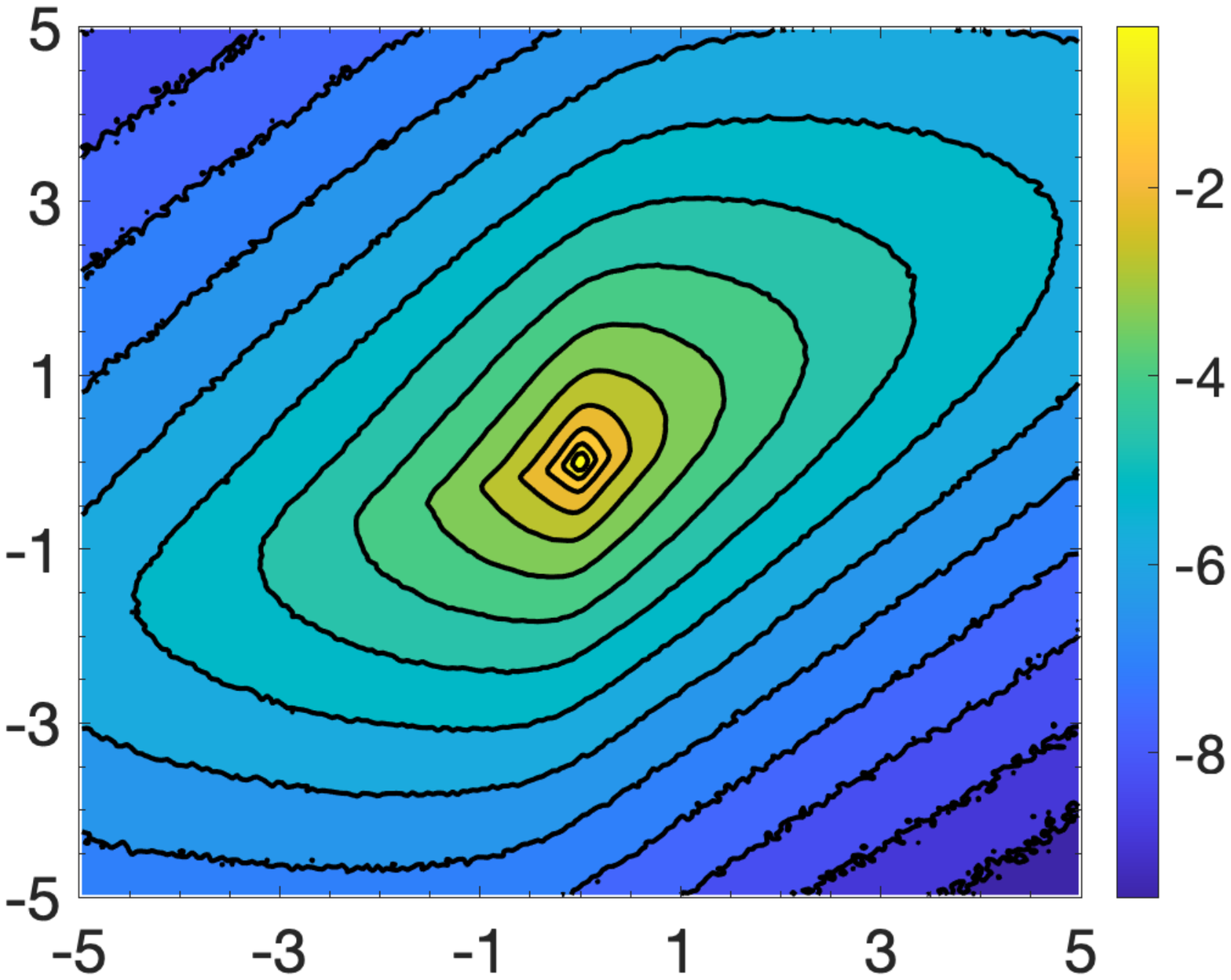}
				\put(80,8){$\Pi_K^F/\langle\Pi_K^F\rangle$}
				\put(5,65){\rotatebox{90}{$\Pi_K/\langle\Pi_K\rangle$}}
		\end{overpic}}
				\subfloat[]
		{\begin{overpic}
				[trim = 0mm 50mm 0mm 70mm,scale=0.32,clip,tics=20]{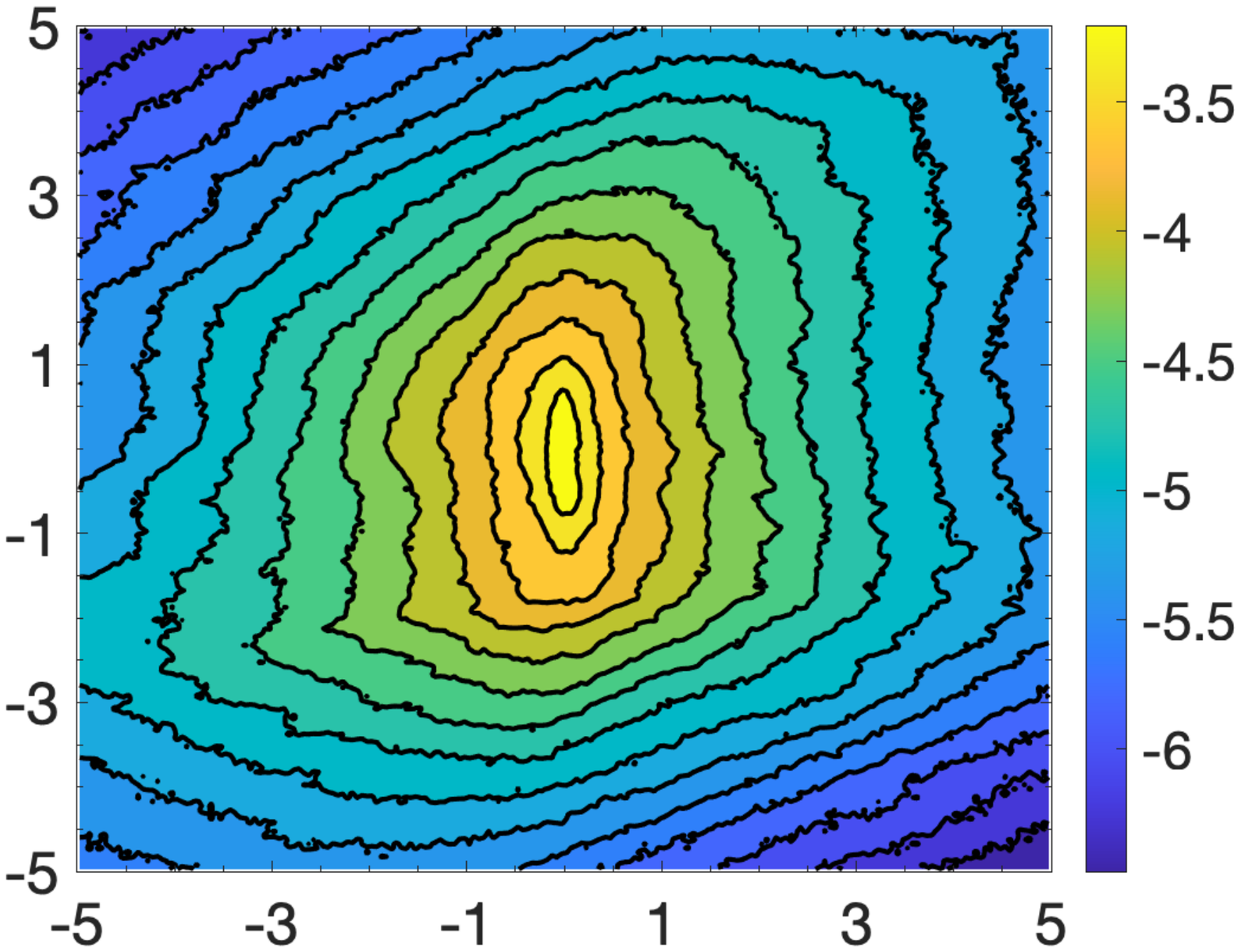}
				\put(80,8){$\Pi_K^F/\langle\Pi_K^F\rangle$}
				\put(5,65){\rotatebox{90}{$\Pi_K/\langle\Pi_K\rangle$}}			
		\end{overpic}}
		\caption{Contour plot of the logarithm of the joint PDF of $\Pi_K/\langle\Pi_K\rangle$ and $\Pi_K^F/\langle\Pi_K^F\rangle$ for (a) $\ell/\eta=0.25$, (b) $\ell/\eta=6$, (c) $\ell/\eta=16$, (d) $\ell/\eta=60$. Colors correspond to the logarithm of the PDF.} 
		\label{JPF_PiK_FSG}
\end{figure}}
In the context of LES, the contributions to $\Pi_K$ and $\Pi_P$ from the filtered field dynamics, namely $\Pi_K^F\equiv \Pi_K^{F,SSA}+\Pi_K^{F,VS}$ and $\Pi_P^F$ (see  \S\ref{FMech}), are closed since they depend only on the filtered field quantities, not the sub-grid fields. It is therefore of particular interest to understand how these terms contribute to the total TKE and TPE fluxes in the flow. To explore this, in figure \ref{JPF_PiK_FSG} we consider the joint PDF of $\Pi_K$ and $\Pi_K^F$ for different filtering lengths $\ell/\eta$. The filtered fluxes satisfy $\lim_{\ell/\eta\to 0}\Pi_K^F\to\Pi_K$, and the results for $\ell/\eta=0.25$ are approaching this regime, showing a very small probability of events deviating from the state $\Pi_K=\Pi_K^F$. As $\ell/\eta$ is increased, there remains a significant positive correlation between the variables  $\Pi_K$ and $\Pi_K^F$, indicating that $\Pi_K^F$ makes an important contribution to the total flux $\Pi_K$, consistent with the mean-field results in figure \ref{TKE_budgets}.
{\vspace{0mm}\begin{figure}
		\centering
		\subfloat[]
		{\begin{overpic}
				[trim = 0mm 50mm 0mm 70mm,scale=0.32,clip,tics=20]{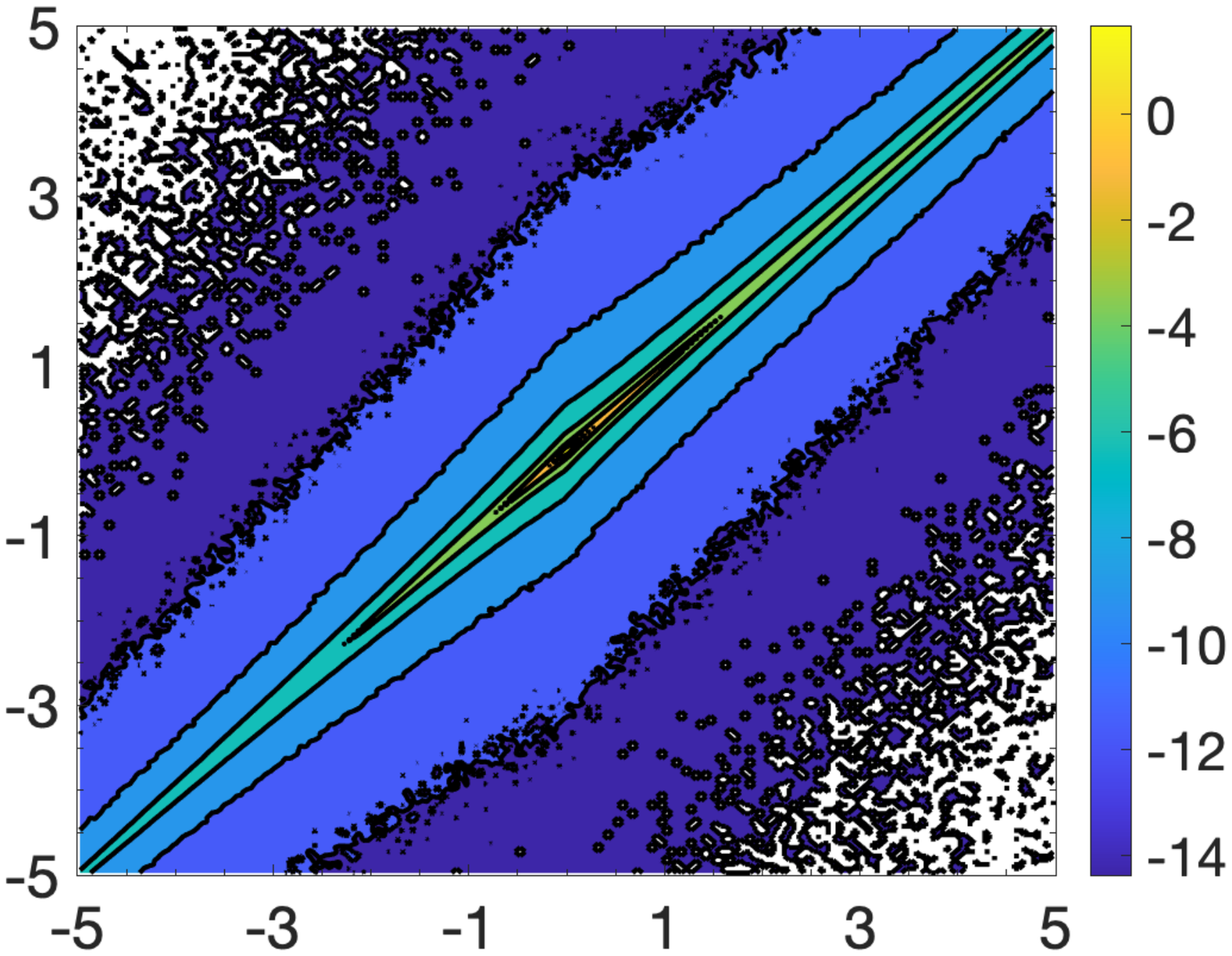}
				\put(80,8){$\Pi_P^F/\langle\Pi_P^F\rangle$}
				\put(5,65){\rotatebox{90}{$\Pi_P/\langle\Pi_P\rangle$}}
		\end{overpic}}
				\subfloat[]
		{\begin{overpic}
				[trim = 0mm 50mm 0mm 70mm,scale=0.32,clip,tics=20]{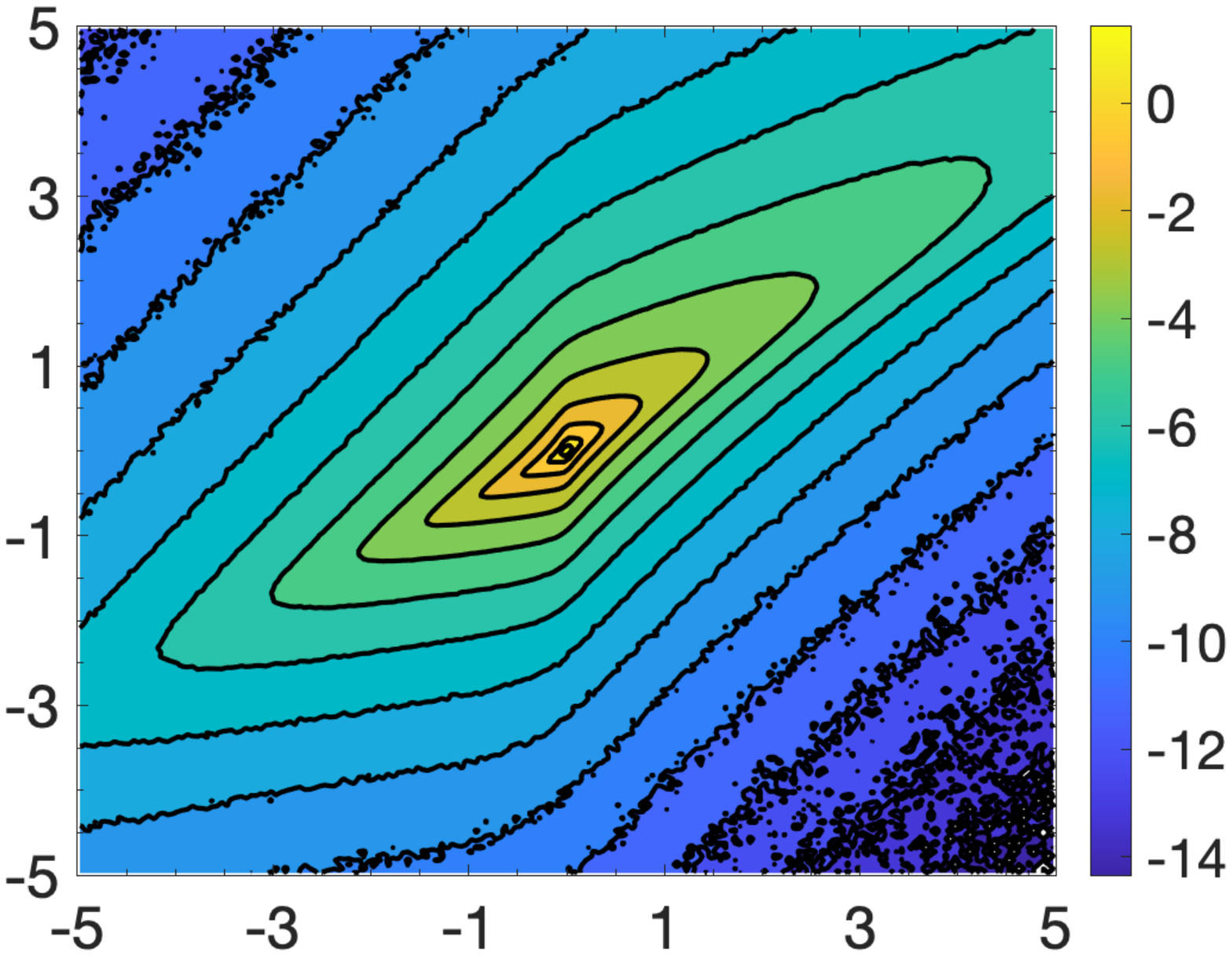}
				\put(80,8){$\Pi_P^F/\langle\Pi_P^F\rangle$}
				\put(5,65){\rotatebox{90}{$\Pi_P/\langle\Pi_P\rangle$}}	
		\end{overpic}}\\
			\subfloat[]
				{\begin{overpic}
				[trim = 0mm 50mm 0mm 70mm,scale=0.32,clip,tics=20]{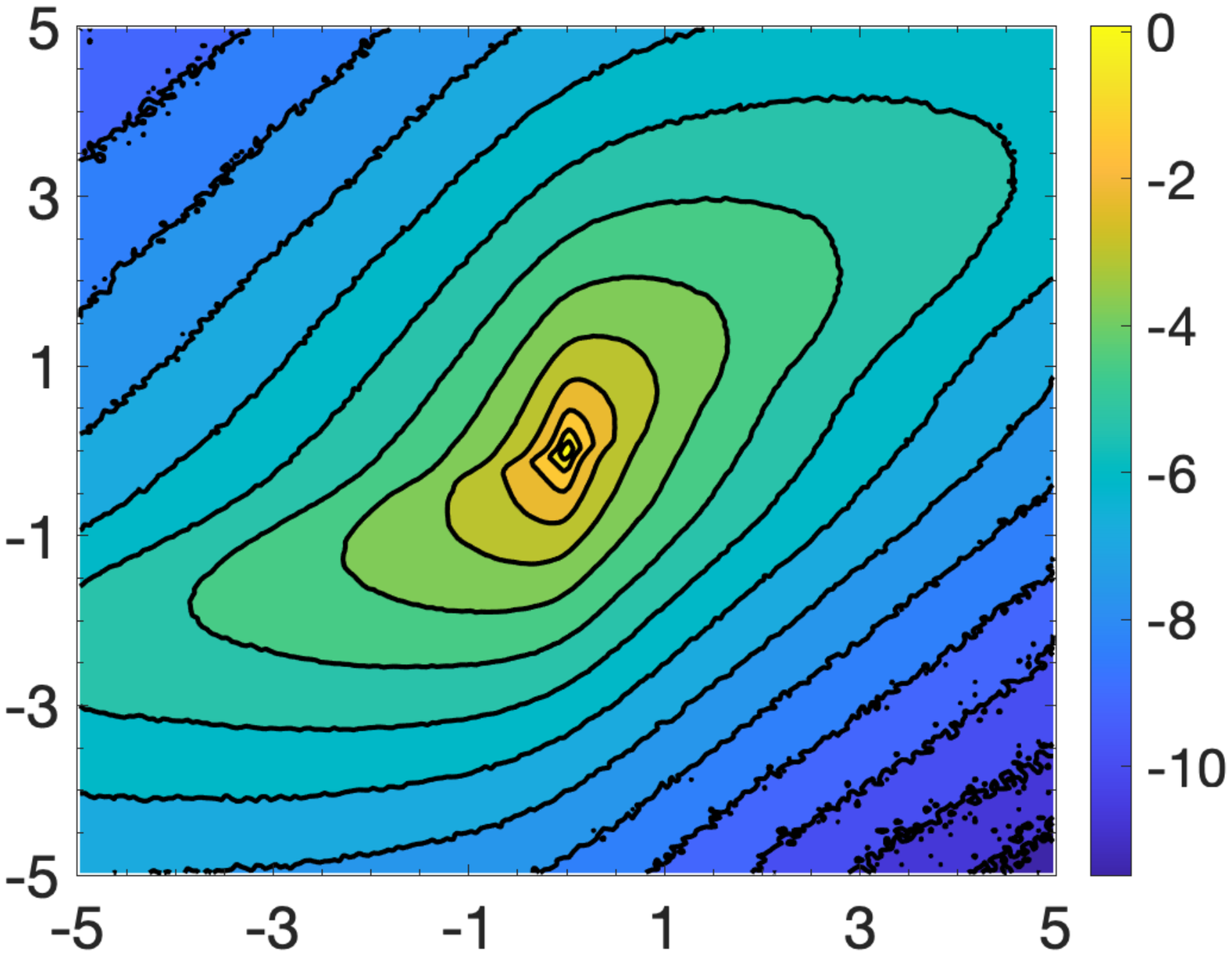}
				\put(80,8){$\Pi_P^F/\langle\Pi_P^F\rangle$}
				\put(5,65){\rotatebox{90}{$\Pi_P/\langle\Pi_P\rangle$}}
		\end{overpic}}
				\subfloat[]
		{\begin{overpic}
				[trim = 0mm 50mm 0mm 70mm,scale=0.32,clip,tics=20]{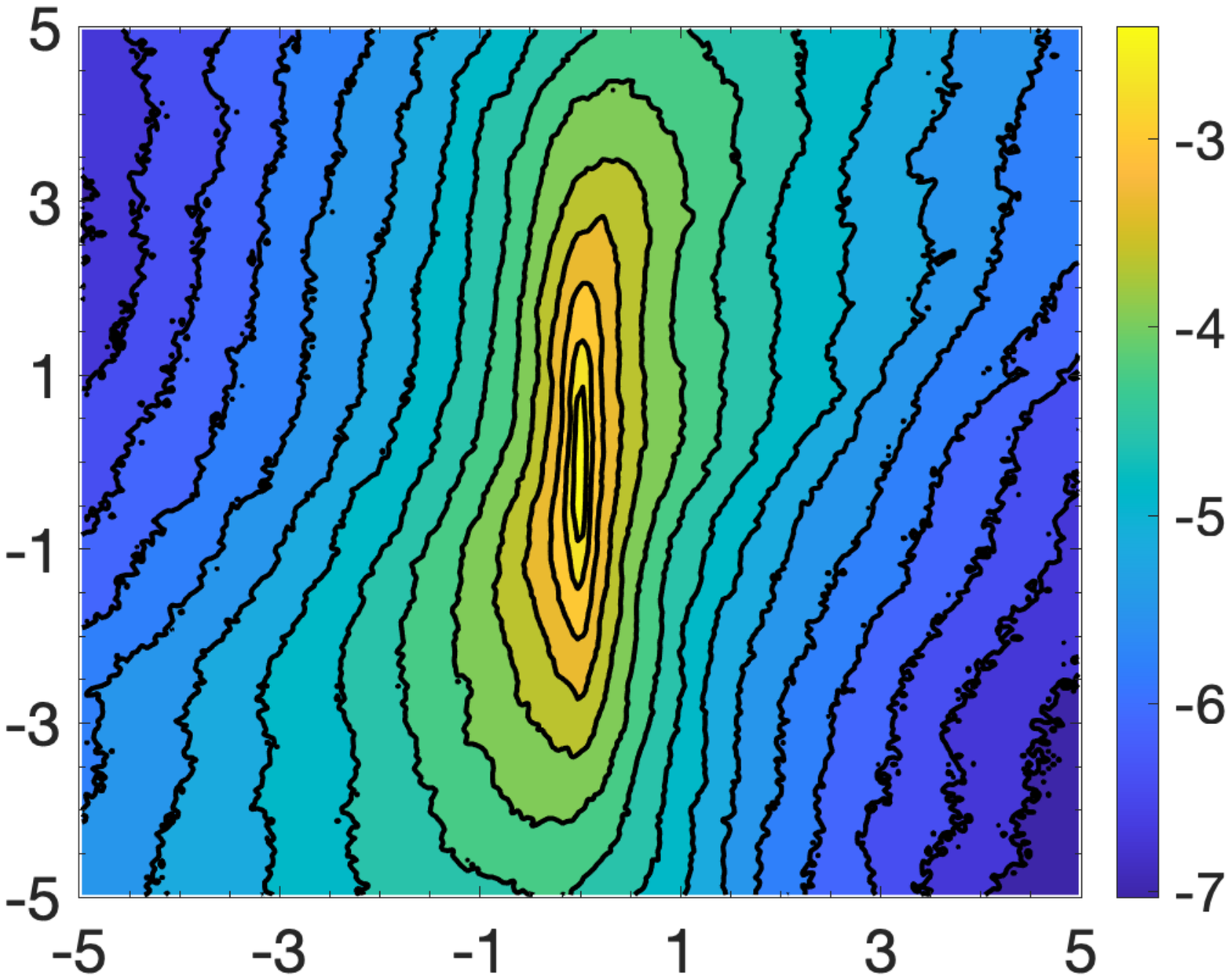}
				\put(80,8){$\Pi_P^F/\langle\Pi_P^F\rangle$}
				\put(5,65){\rotatebox{90}{$\Pi_P/\langle\Pi_P\rangle$}}			
		\end{overpic}}
		\caption{Contour plot of the logarithm of the joint PDF of $\Pi_P/\langle\Pi_P\rangle$ and $\Pi_P^F/\langle\Pi_P^F\rangle$ for (a) $\ell/\eta=0.25$, (b) $\ell/\eta=6$, (c) $\ell/\eta=16$, (d) $\ell/\eta=60$. Colors correspond to the logarithm of the PDF.} 
		\label{JPF_PiP_FSG}
\end{figure}}
However, there is a significant probability of events where either $\Pi_K^F$ and $\Pi_K$ have significantly different magnitudes, or else even have opposite signs. In terms of the physical mechanisms discussed in \S\ref{FMech}, this means that at a given scale $\ell$, if the strain and vorticity fields are being amplified at that scale by nonlinearity such that $\Pi_K^F>0$, this may nevertheless not lead to a downscale flux of TKE associated with $\Pi_K>0$ since at scales smaller than $\ell$ the strain and vorticity fields may be experiencing suppression (rather than amplification) due to nonlinearity yielding $\Pi_K^{SG}<0$, and if this is strong enough then $\Pi_K=\Pi_K^F+\Pi_K^{SG}<0$. The spread of the PDF about the line $\Pi_K=\Pi_K^F$ also implies that the relative contribution of $\Pi_K^F$ to $\Pi_K$ is similar both during events where $\Pi_K\sim \langle \Pi_K\rangle$ and large fluctuations where $|\Pi_K|\ll \langle \Pi_K\rangle$ or $|\Pi_K|\gg \langle \Pi_K\rangle$. For $\ell/\eta=60$, the results show that $\Pi_K$ and $\Pi_K^F$ are almost uncorrelated, and at this scale, $\langle \Pi_K\rangle/\langle\Pi_K^F\rangle\approx 43$, so that the filtered field makes a small, uncorrelated contribution to $\Pi_K$, which is itself very small at $\ell/\eta=60$. The results for the joint PDF of $\Pi_P$ and $\Pi_P^F$ are shown in figure \ref{JPF_PiP_FSG}. These are very similar to the TKE flux results, with $\Pi_P^F\sim \Pi_P$ observed at the smallest scales, and a significant correlation between the two at intermediate scales. The correlation between $\Pi_P^F$ and $\Pi_P$ is, however, slightly stronger than that between $\Pi_K$ and $\Pi_K^F$.

{\vspace{0mm}\begin{figure}
		\centering
		\subfloat[]
		{\begin{overpic}
				[trim = 0mm 50mm 0mm 70mm,scale=0.32,clip,tics=20]{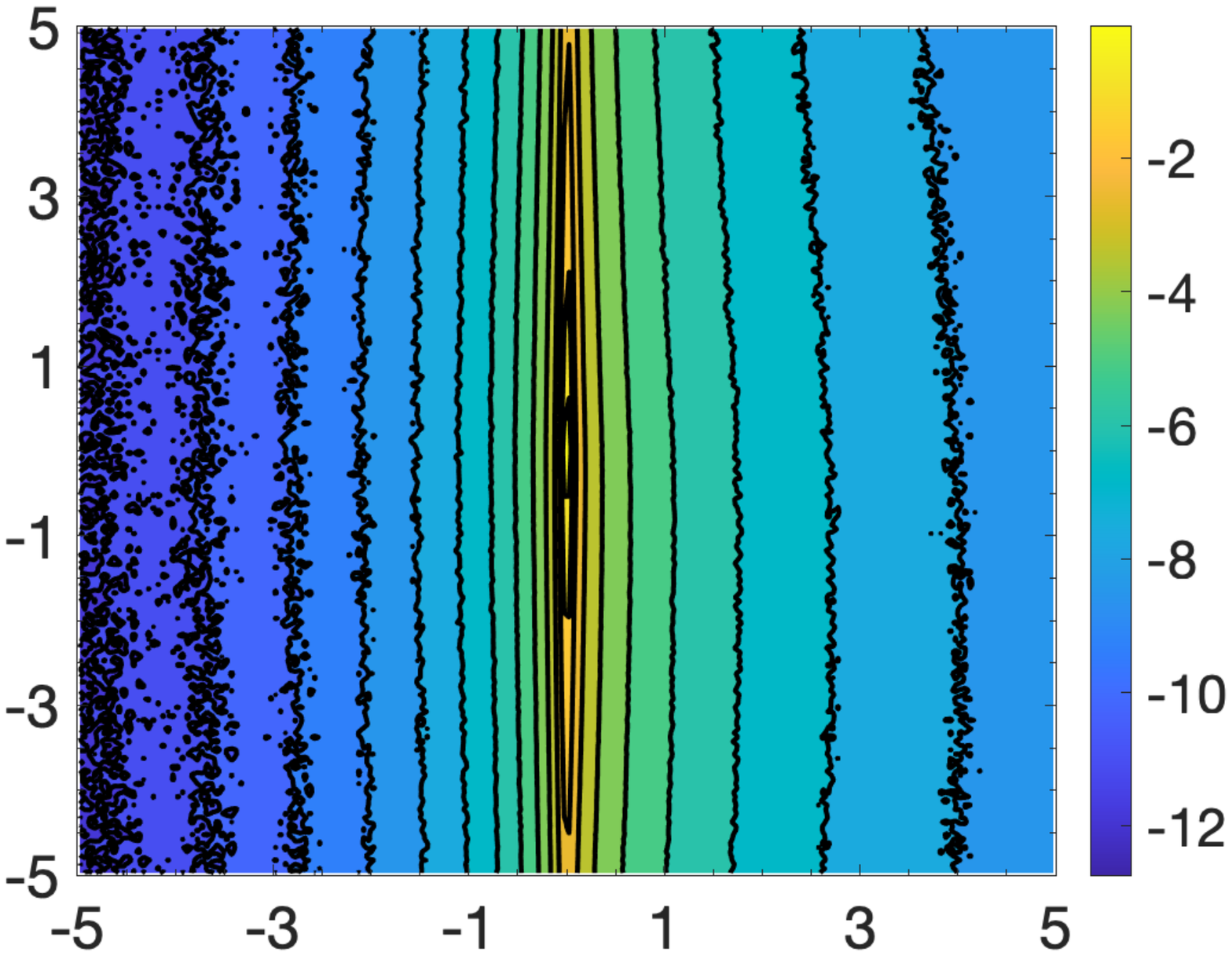}
				\put(5,70){\rotatebox{90}{$\mathcal{B}/\langle\mathcal{B}\rangle$}}
				\put(80,8){$\Pi_K/\langle\Pi_K\rangle$}
		\end{overpic}}
				\subfloat[]
		{\begin{overpic}
				[trim = 0mm 50mm 0mm 70mm,scale=0.32,clip,tics=20]{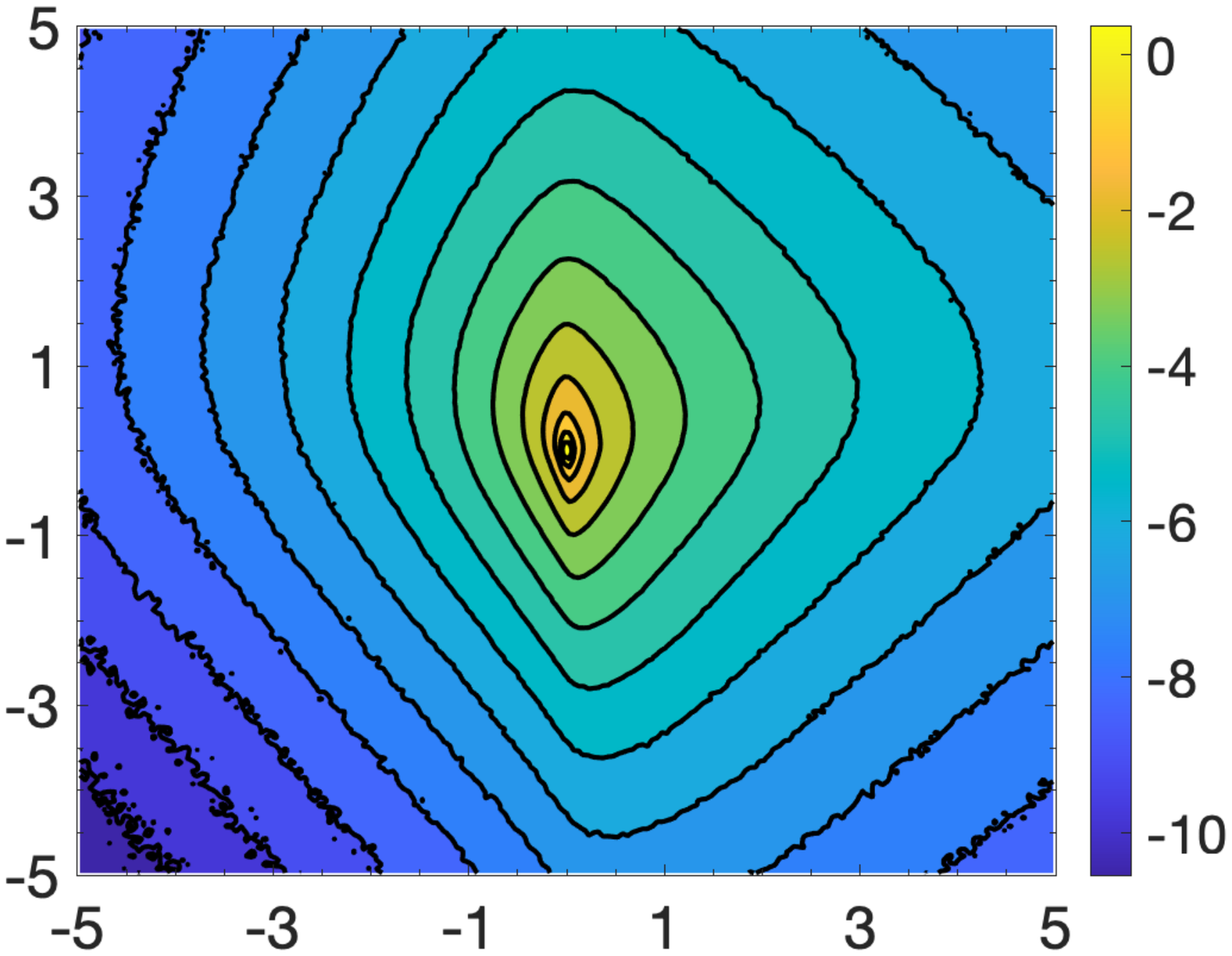}
				\put(5,70){\rotatebox{90}{$\mathcal{B}/\langle\mathcal{B}\rangle$}}
				\put(80,8){$\Pi_K/\langle\Pi_K\rangle$}		
		\end{overpic}}\\
			\subfloat[]
				{\begin{overpic}
				[trim = 0mm 50mm 0mm 70mm,scale=0.32,clip,tics=20]{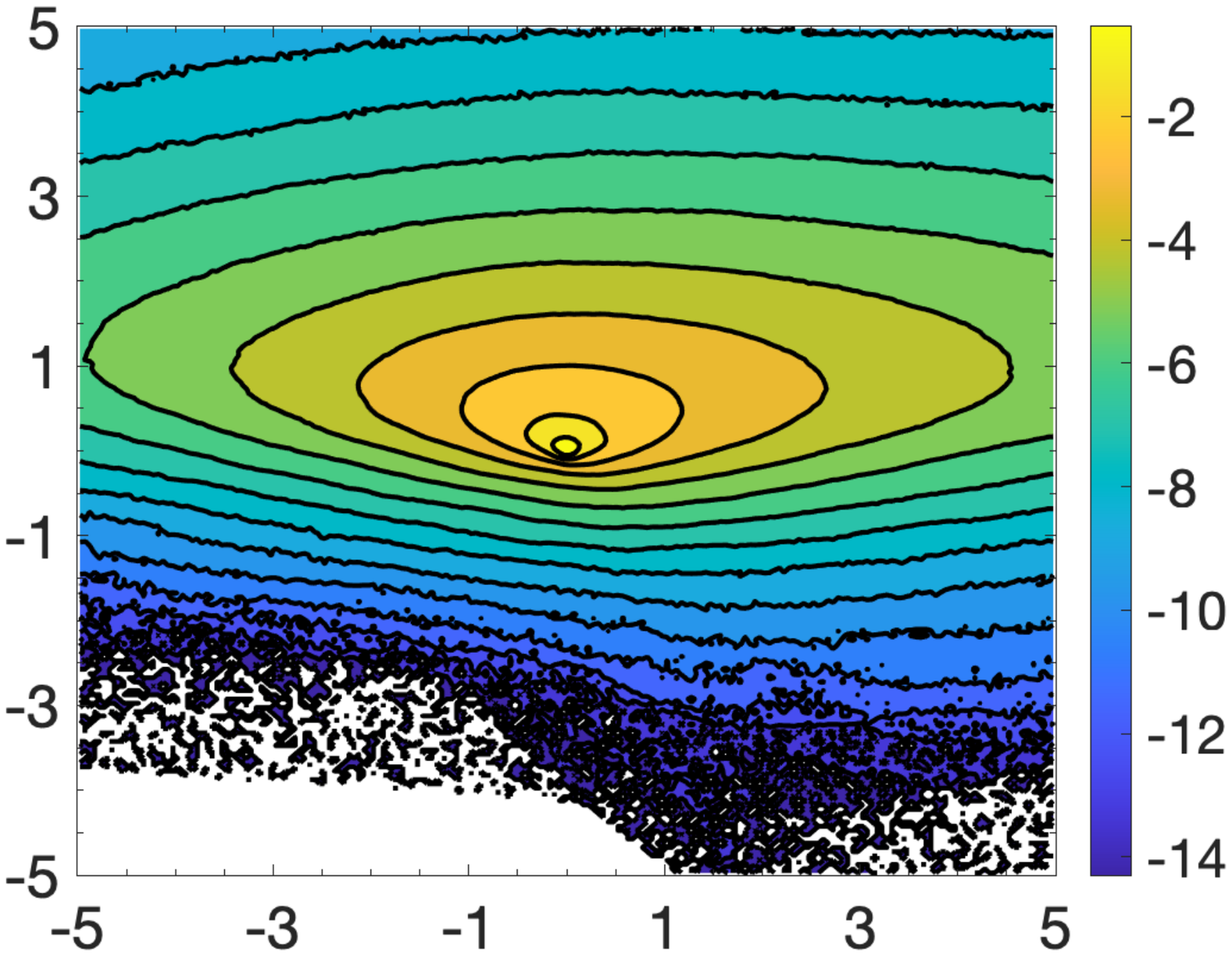}
				\put(5,70){\rotatebox{90}{$\mathcal{B}/\langle\mathcal{B}\rangle$}}
				\put(80,8){$\Pi_K/\langle\Pi_K\rangle$}
		\end{overpic}}
				\subfloat[]
		{\begin{overpic}
				[trim = 0mm 50mm 0mm 70mm,scale=0.32,clip,tics=20]{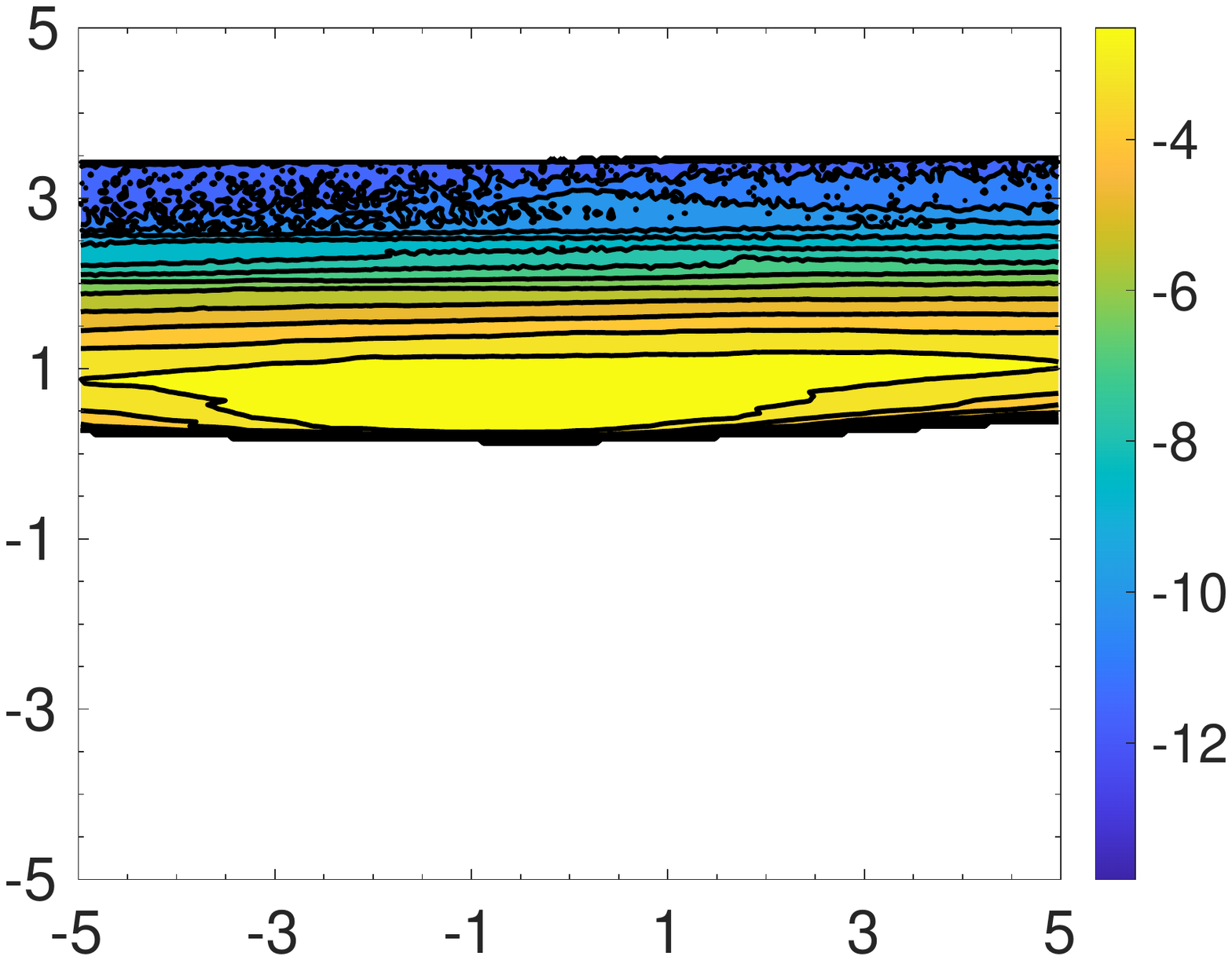}
				\put(5,70){\rotatebox{90}{$\mathcal{B}/\langle\mathcal{B}\rangle$}}
				\put(80,8){$\Pi_K/\langle\Pi_K\rangle$}			
		\end{overpic}}
		\caption{Contour plot of the logarithm of the joint PDF of $\mathcal{B}/\langle \mathcal{B}\rangle$ and $\Pi_K/\langle\Pi_K\rangle$ for (a) $\ell/\eta=0.25$, (b) $\ell/\eta=6$, (c) $\ell/\eta=16$, (d) $\ell/\eta=60$. Colors correspond to the logarithm of the PDF.} 
		\label{JPF_PiK_B}
\end{figure}}
%
In figure \ref{JPF_PiK_B} we consider the joint PDF of $\mathcal{B}$ and $\Pi_K$ for different filter lengths $\ell$. The results show that as $\ell$ is decreased, the PDF reorients from being extended along the $\Pi_K$ axis, to being extended along the $\mathcal{B}$ axis. This corresponds to the transition from larger scales where buoyancy plays a more dominant role in the TKE energetics, to smaller scales where the nonlinear energy flux plays a dominant role. As also observed in figure \ref{PDF_various}(d), at larger scales the probability to observe $\mathcal{B}/\langle\mathcal{B}\rangle<0$ is very low, suggesting that convective motion is very rare at these scales. As $\ell$ decreases, however, the probability to observe $\mathcal{B}/\langle\mathcal{B}\rangle<0$ increases significantly. At $\ell/\eta=16$, $-\langle\mathcal{B}\rangle\approx\langle \Pi_K\rangle$, and the results in figure \ref{JPF_PiK_B} show that at this scale, convective motion $\mathcal{B}/\langle\mathcal{B}\rangle<0$ can occur, but the PDF is strongly skewed towards stably stratified regions that have $\mathcal{B}/\langle\mathcal{B}\rangle>0$. Moreover, for this scale the PDF is stretched significantly along the $\Pi_K$ axis, showing that fluctuations of $\Pi_K$ are considerably stronger than those of $\mathcal{B}$ about their respective mean values which are approximately equal in magnitude at this scale. The results for the joint PDF of $\mathcal{B}$ and $\Pi_P$ are similar to those of the joint PDF of $\mathcal{B}$ and $\Pi_K$, and therefore for brevity we do not show them.
{\vspace{0mm}\begin{figure}
		\centering
		\subfloat[]
		{\begin{overpic}
				[trim = 0mm 50mm 0mm 70mm,scale=0.32,clip,tics=20]{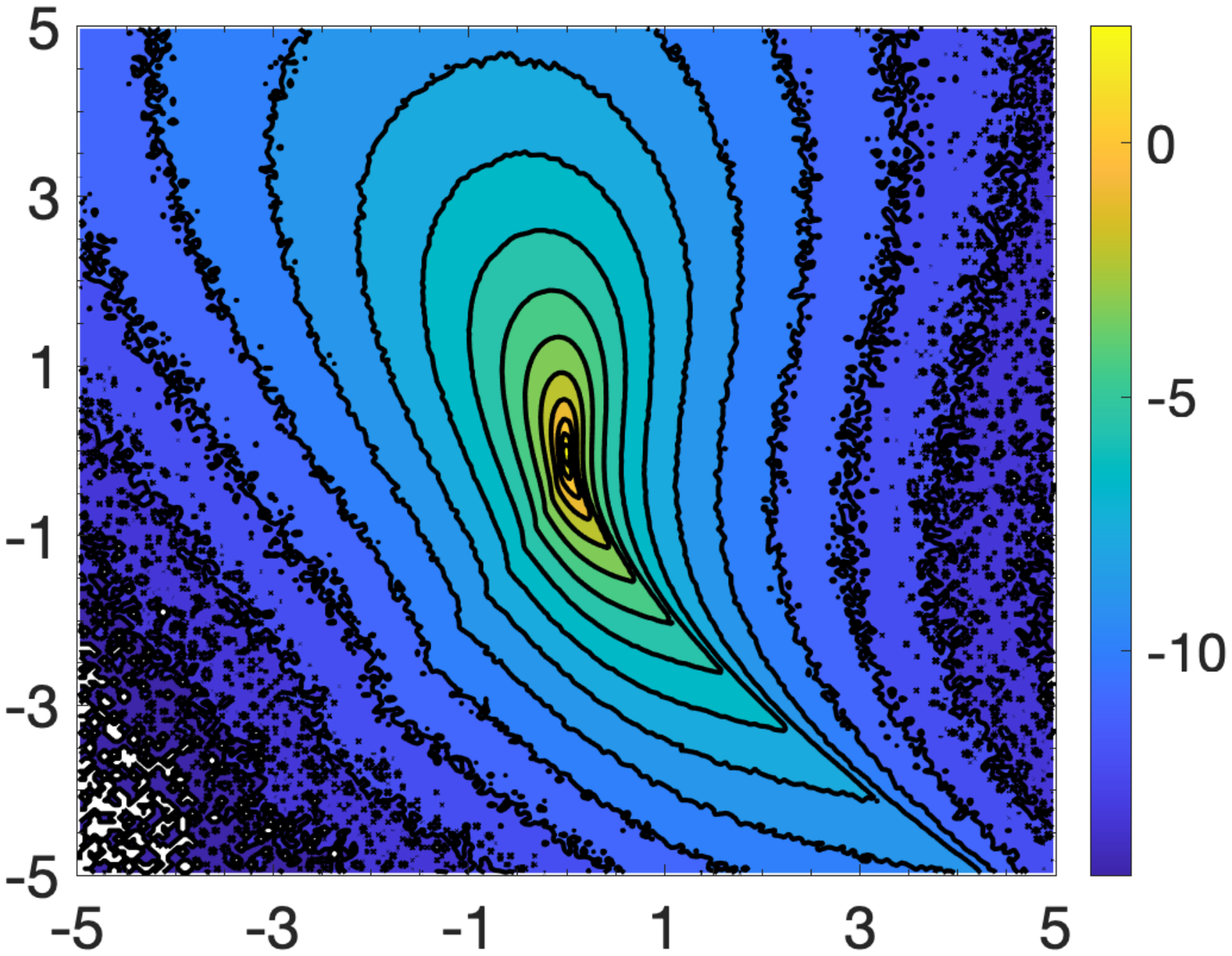}
				\put(80,8){$R/\langle \|\widetilde{\boldsymbol{s}}\|^2\rangle^{3/2}$}
				\put(5,65){\rotatebox{90}{$Q/\langle \|\widetilde{\boldsymbol{s}}\|^2\rangle$}}
		\end{overpic}}
				\subfloat[]
		{\begin{overpic}
				[trim = 0mm 50mm 0mm 70mm,scale=0.32,clip,tics=20]{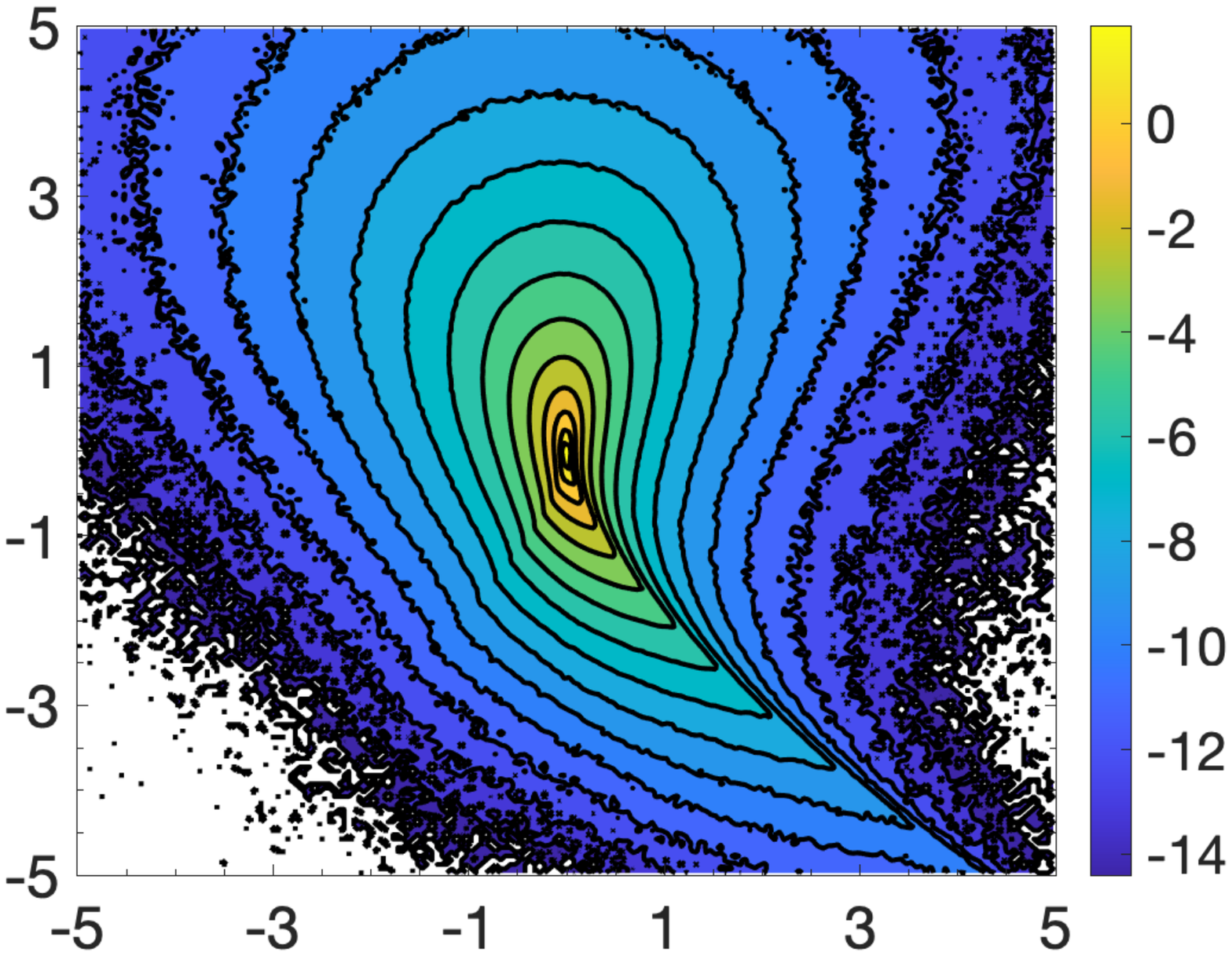}
				\put(80,8){$R/\langle \|\widetilde{\boldsymbol{s}}\|^2\rangle^{3/2}$}
				\put(5,65){\rotatebox{90}{$Q/\langle \|\widetilde{\boldsymbol{s}}\|^2\rangle$}}
		\end{overpic}}\\
			\subfloat[]
				{\begin{overpic}
				[trim = 0mm 50mm 0mm 70mm,scale=0.32,clip,tics=20]{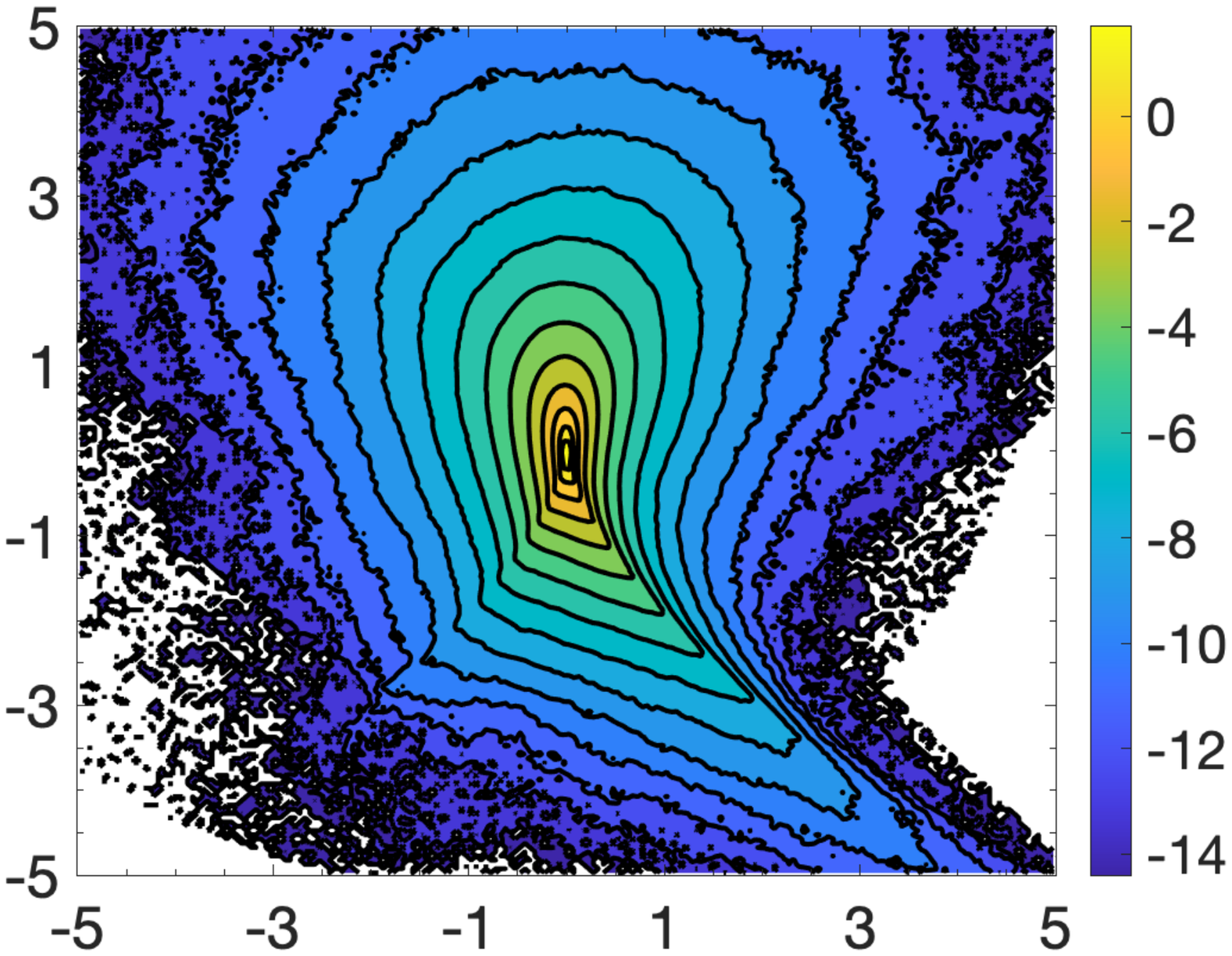}
				\put(80,8){$R/\langle \|\widetilde{\boldsymbol{s}}\|^2\rangle^{3/2}$}
				\put(5,65){\rotatebox{90}{$Q/\langle \|\widetilde{\boldsymbol{s}}\|^2\rangle$}}
		\end{overpic}}
				\subfloat[]
		{\begin{overpic}
				[trim = 0mm 50mm 0mm 70mm,scale=0.32,clip,tics=20]{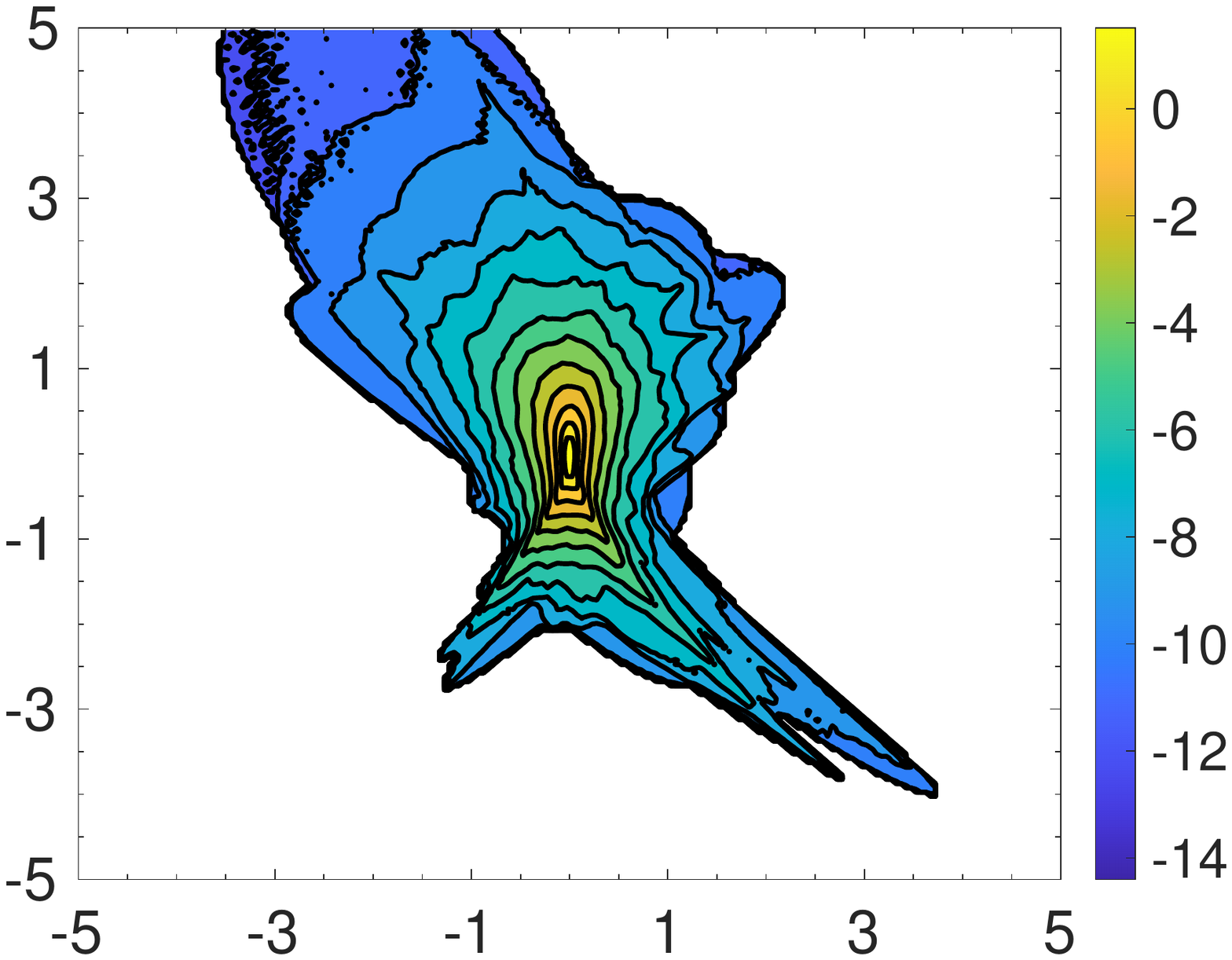}
				\put(80,8){$R/\langle \|\widetilde{\boldsymbol{s}}\|^2\rangle^{3/2}$}
				\put(5,65){\rotatebox{90}{$Q/\langle \|\widetilde{\boldsymbol{s}}\|^2\rangle$}}			
		\end{overpic}}
		\caption{Contour plot of the logarithm of the joint PDF of $Q$ and $R$ for (a) $\ell/\eta=0.25$, (b) $\ell/\eta=6$, (c) $\ell/\eta=16$, (d) $\ell/\eta=60$. Colors correspond to the logarithm of the PDF.} 
		\label{JPF_QR}
\end{figure}}

Finally, given the importance of the velocity gradient dynamics for the TKE flux, in figure \ref{JPF_QR} we plot the results for the joint PDF of the filtered velocity gradient invariants $Q\equiv {\boldsymbol{\nabla\widetilde{u}}\boldsymbol{:}\boldsymbol{\nabla\widetilde{u}}}$ and $R\equiv {(\boldsymbol{\nabla\widetilde{u}}\boldsymbol{\cdot}\boldsymbol{\nabla\widetilde{u}}) \boldsymbol{:}\boldsymbol{\nabla\widetilde{u}}}$ at different filter scales $\ell/\eta$. In \cite{danish18} this joint PDF was considered for isotropic turbulence, and they found that the classic sheared-drop shape of the isoprobability lines is preserved as $\ell/\eta$ is varied, while other statistical characterizations of the velocity gradients also reveal behavior that is qualitatively similar as the filter scale is increased \citep{tom20}. As a result of this, the change in shape of the PDF contours shown in figure \ref{JPF_QR} as  $\ell/\eta$ is increased is strictly due to the competing influence of buoyancy and mean-shear with nonlinearity in the flow. For $\ell/\eta=0.25$, the shape of the PDF is identical to that observed in isotropic turbulence \citep{meneveau11}, with the contours extending down the right Viellefosse tail, and a strong preference for quadrants $Q>0, R<0$ and $Q<0, R>0$. As $\ell/\eta$ is increased, however, the PDF becomes more symmetric in $R$. This is due to the fact that as $\ell/\eta$ is increased, the nonlinear term that generates the asymmetry in $R$ (which is associated with the preference for vortex stretching over compression, and strain-rate self-amplification over self-suppression \citep{tsinober}) becomes subleading compared to buoyancy and mean-shear, and these terms, being linear, generate time-reversible dynamics with no preference for nonlinear amplification or suppression of the strain-rate and vorticity fields. Figure \ref{JPF_QR} (d) still shows, however, that for the outer isoprobability lines, an asymmetry still exists, with a preference for the regions  $Q>0, R<0$ and $Q<0, R>0$. This corresponds to events in the flow where locally the nonlinear term dominates in the flow over buoyancy, allowing for strong strain-rate and/or vorticity amplification in that localized region.

\section{Conclusions}\label{Conc}

In this paper, we have analyzed the scale-dependent TKE and TPE in sheared, stably stratified turbulence (SSST) using a filtering approach, where the flow has constant mean velocity gradient and mean density gradient. Equations for the scale-dependent TKE and TPE are to explore the competing effects in the flow as well as the physical mechanisms governing the TKE and TPE fluxes between scales. Various quantities in these equations were then evaluated using data from DNS of SSST, with attention given to both the mean field behavior of the flow, as well as fluctuations about this mean field state.

In terms of the mean field properties, while the mean TKE $\langle e_K\rangle$ is larger than the mean TPE $\langle e_P\rangle$ by an order of magnitude at the large scales, the difference between them reduces with reducing scale $\ell$. The TPE nevertheless plays an import energetic role in the system, and is larger than the vertical component of TKE at all scales in the flow. The mean small-scale TKE dissipation rate $\langle \varepsilon_K\rangle$ is also significantly larger than the potential dissipation rate $\langle \varepsilon_P\rangle$. The mean TKE and TPE fluxes between scales, $\langle\Pi_K\rangle$ and $\langle\Pi_P\rangle$, respectively, do not reveal a cascade regime in the flow due to the impact of the mean-shear down to relatively small-scales in the flow. The contributions to these fluxes from the filtered fields are also shown the be significant at scales where the energy fluxes play significant roles in the energy budget equations, however, the contribution from the sub-grid fields are also import, just as has been observed for isotropic turbulence.

To understand the flow energetics beyond the mean fields, PDFs of various quantities (normalized by their mean values) have been studied and discussed. The PDFs of small-scale TKE and TPE are highly non-Gaussian at the smallest scales, indicating large fluctuations of the TKE and TPE about their mean field values. The TPE shows stronger fluctuations from its mean field value than the TKE, which is consistent with the known result that scalar fields are more intermittent the velocity fields in turbulence, since scalar fields lack a pressure gradient term in their equation which can act to suppress large fluctuations. As the filter scale is increased, these PDFs approach a Gaussian distribution at scales where the linear forces in the flow (mean-shear and buoyancy) dominate the dynamics. At larger scales in the flow, buoyancy seems always convert TKE to TPE, with the PDF of the buoyancy term $\mathcal{B}$ showing almost zero probability of locally convective events where $\mathcal{B}>0$. The probability of locally convective regions increases as the filter scale is decreased, however, the probability remains small at scales downward to the Ozimidov scale. The TKE and TPE fluxes between scales are on average positive (downscale flux), with their instantaneous values positively correlated, and increasingly so with decreasing filter scale. However, the correlation is quite weak and this is because the physical mechanisms governing the TKE and TPE fluxes are quite different, as discussed in our theoretical analysis in \S\ref{FMech}. Indeed, the joint PDFs of the TKE and TPE fluxes reveals a significant probability of events where the TKE and TPE fluxes may have very different magnitudes, and even opposite signs. Finally, the PDF of the principal invariants of the filtered velocity gradients (the so-called $Q,R$ invariants \citep{tsinober}) reveals the classical sheared-drop shape for the smallest filter scales. However, the PDF becomes increasingly symmetric about $R=0$ as the filter scale increases. This contrasts with the behavior observed in isotropic turbulence \citep{danish18} and is due to the increasingly dominant effects of the linear mean shear and buoyancy effects over the nonlinear inertial effects at these scales. This symmetry that emerges means that vortex stretching and compression become equally probably, as well as strain self-amplification and self-suppression.

\section{Acknowledgements}
A.D. Bragg and X. Zhang acknowledge support through a National Science Foundation (NSF) CAREER award \# 2042346. This work used the Extreme Science and Engineering Discovery Environment (XSEDE), which is supported by NSF grant number ACI-1548562 \citep{xsede}. Specifically, the Comet and Expanse clusters were used under allocation
CTS170009.  The DNS database was generated with funding from the U.S.\ Office of Naval Research via grant
N00014-15-1-2248.  High performance computing resources for the DNS were provided through
the U.S.\ Department of Defense High Performance Computing Modernization
Program by the Army Engineer Research and Development Center and the Army
Research Laboratory under Frontier Project FP-CFD-FY14-007.

This work was performed under the auspices of the U.S. Department of Energy by Lawrence Livermore National Laboratory under contract DE-AC52-07NA27344.

\section{Declaration of Interests}

The authors report no conflict of interest.

\bibliographystyle{jfm}
\bibliography{SSST_JFM}

\begin{thebibliography}{54}
\expandafter\ifx\csname natexlab\endcsname\relax\def\natexlab#1{#1}\fi
\def\au#1{#1} \def\ed#1{#1} \def\yr#1{#1}\def\at#1{#1}\def\jt#1{\textit{#1}}
  \def\bt#1{#1}\def\bvol#1{\textbf{#1}} \def\vol#1{#1} \def\pg#1{#1}
  \def\publ#1{#1}\def\arxiv#1{#1}\def\org#1{#1}\def\st#1{\textit{#1}}

\bibitem[Alam {\em et~al.\/}(2019)Alam, Guha \& Verma]{alam19}
{\sc \au{Alam, Shadab}, \au{Guha, Anirban} \& \au{Verma, Mahendra~K.}}
  \yr{2019}  \at{Revisiting bolgiano–obukhov scaling for moderately stably
  stratified turbulence}.  \jt{Journal of Fluid Mechanics}  \bvol{875},
  \pg{961–973}.

\bibitem[Almalkie \& de~Bruyn~Kops(2012)]{almalkie12}
{\sc \au{Almalkie, Saba} \& \au{de~Bruyn~Kops, Stephen~M.}} \yr{2012}
  \at{Kinetic energy dynamics in forced, homogeneous, and axisymmetric stably
  stratified turbulence}.  \jt{Journal of Turbulence}  \bvol{13},  \pg{N29},
  \arxiv{arXiv: https://doi.org/10.1080/14685248.2012.702909}.

\bibitem[Aluie \& Eyink(2009)]{aluie09}
{\sc \au{Aluie, Hussein} \& \au{Eyink, Gregory~L.}} \yr{2009}  \at{Localness of
  energy cascade in hydrodynamic turbulence. ii. sharp spectral filter}.
  \jt{Physics of Fluids}  \bvol{21}~(11),  \pg{115108},  \arxiv{arXiv:
  https://doi.org/10.1063/1.3266948}.

\bibitem[Ayet {\em et~al.\/}(2020)Ayet, Katul, Bragg \& Redelsperger]{ayet20}
{\sc \au{Ayet, A.}, \au{Katul, G.~G.}, \au{Bragg, A.~D.} \& \au{Redelsperger,
  J.~L.}} \yr{2020}  \at{Scalewise return to isotropy in stratified boundary
  layer flows}.  \jt{Journal of Geophysical Research: Atmospheres}
  \bvol{125}~(16),  \pg{e2020JD032732}, e2020JD032732 10.1029/2020JD032732,
  \arxiv{arXiv:
  https://agupubs.onlinelibrary.wiley.com/doi/pdf/10.1029/2020JD032732}.

\bibitem[Betchov(1956)]{betchov56}
{\sc \au{Betchov, R.}} \yr{1956}  \at{An inequality concerning the production
  of vorticty in isotropic turbulence}.  \jt{J. Fluid Mech.}  \bvol{1},
  \pg{497--504}.

\bibitem[Billant \& Chomaz(2001)]{billant01}
{\sc \au{Billant, Paul} \& \au{Chomaz, Jean-Marc}} \yr{2001}
  \at{Self-similarity of strongly stratified inviscid flows}.  \jt{Physics of
  Fluids}  \bvol{13}~(6),  \pg{1645--1651},  \arxiv{arXiv:
  https://doi.org/10.1063/1.1369125}.

\bibitem[Bolgiano~Jr.(1959)]{bolgiano59}
{\sc \au{Bolgiano~Jr., R.}} \yr{1959}  \at{Turbulent spectra in a stably
  stratified atmosphere}.  \jt{Journal of Geophysical Research (1896-1977)}
  \bvol{64}~(12),  \pg{2226--2229},  \arxiv{arXiv:
  https://agupubs.onlinelibrary.wiley.com/doi/pdf/10.1029/JZ064i012p02226}.

\bibitem[Brethouwer {\em et~al.\/}(2007)Brethouwer, Billant, Lindborg \&
  Chomaz]{brethouwer07}
{\sc \au{Brethouwer, G.}, \au{Billant, P.}, \au{Lindborg, E.} \& \au{Chomaz,
  J.-M.}} \yr{2007}  \at{Scaling analysis and simulation of strongly stratified
  turbulent flows}.  \jt{Journal of Fluid Mechanics}  \bvol{585},
  \pg{343?368}.

\bibitem[Brucker {\em et~al.\/}(2007)Brucker, Isaza, Vaithianathan \&
  Collins]{brucker07}
{\sc \au{Brucker, K.~A.}, \au{Isaza, J.~C.}, \au{Vaithianathan, T.} \&
  \au{Collins, L.~R.}} \yr{2007}  \at{Efficient algorithm for simulating
  homogeneous turbulent shear flow without remeshing}.  \jt{J. Comp. Phys.}
  \bvol{225},  \pg{20--32}.

\bibitem[de~Bruyn~Kops(2015)]{debk15}
{\sc \au{de~Bruyn~Kops, S.~M.}} \yr{2015}  \at{Classical turbulence scaling and
  intermittency in stably stratified {B}oussinesq turbulence}.  \jt{J. Fluid
  Mech.}  \bvol{775},  \pg{436--463}.

\bibitem[Buaria {\em et~al.\/}(2020)Buaria, Bodenschatz \& Pumir]{buaria20}
{\sc \au{Buaria, Dhawal}, \au{Bodenschatz, Eberhard} \& \au{Pumir, Alain}}
  \yr{2020}  \at{Vortex stretching and enstrophy production in high reynolds
  number turbulence}.  \jt{Phys. Rev. Fluids}  \bvol{5},  \pg{104602}.

\bibitem[Carbone \& Bragg(2020)]{carbone20}
{\sc \au{Carbone, M.} \& \au{Bragg, A.~D.}} \yr{2020}  \at{Is vortex stretching
  the main cause of the turbulent energy cascade?}  \jt{Journal of Fluid
  Mechanics}  \bvol{883},  \pg{R2}.

\bibitem[Chung \& Matheou(2012)]{chung12}
{\sc \au{Chung, D.} \& \au{Matheou, G.}} \yr{2012}  \at{Direct numerical
  simulation of stationary homogeneous stratified sheared turbulence}.
  \jt{Journal of Fluid Mechanics}  \bvol{696},  \pg{434–467}.

\bibitem[Danish \& Meneveau(2018)]{danish18}
{\sc \au{Danish, Mohammad} \& \au{Meneveau, Charles}} \yr{2018}  \at{Multiscale
  analysis of the invariants of the velocity gradient tensor in isotropic
  turbulence}.  \jt{Phys. Rev. Fluids}  \bvol{3},  \pg{044604}.

\bibitem[Davidson(2004)]{davidson}
{\sc \au{Davidson, P.~A.}} \yr{2004} {\em Turbulence: an introduction for
  scientists and engineers\/}.  \publ{Oxford}.

\bibitem[Doan {\em et~al.\/}(2018)Doan, Swaminathan, Davidson \&
  Tanahashi]{doan18}
{\sc \au{Doan, N. A.~K.}, \au{Swaminathan, N.}, \au{Davidson, P.~A.} \&
  \au{Tanahashi, M.}} \yr{2018}  \at{Scale locality of the energy cascade using
  real space quantities}.  \jt{Phys. Rev. Fluids}  \bvol{3},  \pg{084601}.

\bibitem[Ferrari \& Wunsch(2009)]{ferrari09}
{\sc \au{Ferrari, Raffaele} \& \au{Wunsch, Carl}} \yr{2009}  \at{Ocean
  circulation kinetic energy: Reservoirs, sources, and sinks}.  \jt{Annual
  Review of Fluid Mechanics}  \bvol{41}~(1),  \pg{253--282}.

\bibitem[Gage(1979)]{gage79}
{\sc \au{Gage, K.~S.}} \yr{1979}  \at{Evidence far a k?5/3 law inertial range
  in mesoscale two-dimensional turbulence}.  \jt{Journal of the Atmospheric
  Sciences}  \bvol{36}~(10),  \pg{1950--1954}.

\bibitem[Germano(1992)]{germano92}
{\sc \au{Germano, M.}} \yr{1992}  \at{Turbulence: the filtering approach}.
  \jt{Journal of Fluid Mechanics}  \bvol{238},  \pg{325–336}.

\bibitem[Gregg {\em et~al.\/}(2018)Gregg, D'Asaro, Riley \& Kunze]{Gregg18}
{\sc \au{Gregg, M.C.}, \au{D'Asaro, E.A.}, \au{Riley, J.J.} \& \au{Kunze, E.}}
  \yr{2018}  \at{Mixing efficiency in the ocean}.  \jt{Annual Review of Marine
  Science}  \bvol{10}~(1),  \pg{443--473}, pMID: 28934598.

\bibitem[Jayne(2009)]{Jayne09}
{\sc \au{Jayne, Steven~R.}} \yr{2009}  \at{{The Impact of Abyssal Mixing
  Parameterizations in an Ocean General Circulation Model}}.  \jt{Journal of
  Physical Oceanography}  \bvol{39}~(7),  \pg{1756--1775}.

\bibitem[Johnson(2020)]{johnson20}
{\sc \au{Johnson, Perry~L.}} \yr{2020}  \at{Energy transfer from large to small
  scales in turbulence by multiscale nonlinear strain and vorticity
  interactions}.  \jt{Phys. Rev. Lett.}  \bvol{124},  \pg{104501}.

\bibitem[Johnson(2021)]{johnson21}
{\sc \au{Johnson, Perry~L.}} \yr{2021}  \at{On the role of vorticity stretching
  and strain self-amplification in the turbulence energy cascade}.  \jt{Journal
  of Fluid Mechanics}  \bvol{922},  \pg{A3}.

\bibitem[Katul {\em et~al.\/}(2013)Katul, Porporato, Manes \&
  Meneveau]{katul13}
{\sc \au{Katul, Gabriel~G.}, \au{Porporato, Amilcare}, \au{Manes, Costantino}
  \& \au{Meneveau, Charles}} \yr{2013}  \at{Co-spectrum and mean velocity in
  turbulent boundary layers}.  \jt{Physics of Fluids}  \bvol{25}~(9),
  \pg{091702},  \arxiv{arXiv: https://doi.org/10.1063/1.4821997}.

\bibitem[Khani \& Waite(2015)]{khani15}
{\sc \au{Khani, Sina} \& \au{Waite, Michael~L.}} \yr{2015}  \at{Large eddy
  simulations of stratified turbulence: the dynamic smagorinsky model}.
  \jt{Journal of Fluid Mechanics}  \bvol{773},  \pg{327–344}.

\bibitem[Kolmogorov(1941)]{kolmogorov41a}
{\sc \au{Kolmogorov, A.~N.}} \yr{1941}  \at{The local structure of turbulence
  in an incompressible viscous fluid for very large {R}eynolds numbers}.
  \jt{Dokl. Akad. Nauk. SSSR}  \bvol{30},  \pg{299--303}.

\bibitem[Kumar {\em et~al.\/}(2014)Kumar, Chatterjee \& Verma]{kumar14}
{\sc \au{Kumar, Abhishek}, \au{Chatterjee, Anando~G.} \& \au{Verma,
  Mahendra~K.}} \yr{2014}  \at{Energy spectrum of buoyancy-driven turbulence}.
  \jt{Phys. Rev. E}  \bvol{90},  \pg{023016}.

\bibitem[Lee {\em et~al.\/}(1990)Lee, Kim \& Moin]{lee90}
{\sc \au{Lee, Moon~Joo}, \au{Kim, John} \& \au{Moin, Parviz}} \yr{1990}
  \at{Structure of turbulence at high shear rate}.  \jt{Journal of Fluid
  Mechanics}  \bvol{216},  \pg{561?583}.

\bibitem[Lilly(1983)]{lilly83}
{\sc \au{Lilly, D.~K.}} \yr{1983}  \at{Stratified turbulence and the mesoscale
  variability of the atmosphere}.  \jt{Journal of the Atmospheric Sciences}
  \bvol{40}~(3),  \pg{749--761}.

\bibitem[Lindborg(2005)]{lindborg05}
{\sc \au{Lindborg, Erik}} \yr{2005}  \at{The effect of rotation on the
  mesoscale energy cascade in the free atmosphere}.  \jt{Geophysical Research
  Letters}  \bvol{32}~(1).

\bibitem[Lindborg(2006)]{lindborg06}
{\sc \au{Lindborg, Erik}} \yr{2006}  \at{The energy cascade in a strongly
  stratified fluid}.  \jt{Journal of Fluid Mechanics}  \bvol{550},
  \pg{207?242}.

\bibitem[L'vov \& Falkovich(1992)]{lvov92}
{\sc \au{L'vov, Victor} \& \au{Falkovich, Gregory}} \yr{1992}
  \at{Counterbalanced interaction locality of developed hydrodynamic
  turbulence}.  \jt{Phys. Rev. A}  \bvol{46},  \pg{4762--4772}.

\bibitem[Meneveau(2011)]{meneveau11}
{\sc \au{Meneveau, Charles}} \yr{2011}  \at{Lagrangian dynamics and models of
  the velocity gradient tensor in turbulent flows}.  \jt{Annual Review of Fluid
  Mechanics}  \bvol{43}~(1),  \pg{219--245}.

\bibitem[Obukhov(1959)]{obukhov59}
{\sc \au{Obukhov, A}} \yr{1959} Effect of archimedean forces on the structure
  of the temperature field in a turbulent flow.  \bt{In {\em Dokl. Akad. Nauk
  SSSR\/}}, ,  \vol{vol. 125},  \pg{pp. 1246--1248}.

\bibitem[Peltier \& Caulfield(2003)]{Peltier03}
{\sc \au{Peltier, W.~R.} \& \au{Caulfield, C.~P.}} \yr{2003}  \at{Mixing
  efficiency in stratified shear flows}.  \jt{Annual Review of Fluid Mechanics}
   \bvol{35}~(1),  \pg{135--167}.

\bibitem[Pope(2000)]{pope}
{\sc \au{Pope, S.~B.}} \yr{2000} {\em Turbulent Flows\/}.  \publ{New York:
  Cambridge University Press}.

\bibitem[Portwood {\em et~al.\/}(2019)Portwood, de~Bruyn~Kops \&
  Caulfield]{portwood19}
{\sc \au{Portwood, G.~D.}, \au{de~Bruyn~Kops, S.~M.} \& \au{Caulfield, C.~P.}}
  \yr{2019}  \at{Asymptotic dynamics of high dynamic range stratified
  turbulence}.  \jt{Phys. Rev. Lett.}  \bvol{122},  \pg{194504}.

\bibitem[Riley \& de~Bruyn~Kops(2003)]{riley03}
{\sc \au{Riley, James~J.} \& \au{de~Bruyn~Kops, Stephen~M.}} \yr{2003}
  \at{Dynamics of turbulence strongly influenced by buoyancy}.  \jt{Physics of
  Fluids}  \bvol{15}~(7),  \pg{2047--2059}.

\bibitem[Riley \& Lelong(2000)]{riley00}
{\sc \au{Riley, James~J.} \& \au{Lelong, Marie-Pascale}} \yr{2000}  \at{Fluid
  motions in the presence of strong stable stratification}.  \jt{Annual Review
  of Fluid Mechanics}  \bvol{32}~(1),  \pg{613--657}.

\bibitem[Riley \& Lindborg(2012)]{riley12}
{\sc \au{Riley, James~J.} \& \au{Lindborg, Erik}} \yr{2012} {\em Recent
  Progress in Stratified Turbulence\/},  \pg{p. 269–317}.  \publ{Cambridge
  University Press}.

\bibitem[Sekimoto {\em et~al.\/}(2016)Sekimoto, Dong \& Jiménez]{sekimoto16}
{\sc \au{Sekimoto, Atsushi}, \au{Dong, Siwei} \& \au{Jiménez, Javier}}
  \yr{2016}  \at{Direct numerical simulation of statistically stationary and
  homogeneous shear turbulence and its relation to other shear flows}.
  \jt{Physics of Fluids}  \bvol{28}~(3),  \pg{035101},  \arxiv{arXiv:
  https://doi.org/10.1063/1.4942496}.

\bibitem[Sujovolsky \& Mininni(2020)]{Sujovolsky20}
{\sc \au{Sujovolsky, N.~E.} \& \au{Mininni, P.~D.}} \yr{2020}  \at{From waves
  to convection and back again: The phase space of stably stratified
  turbulence}.  \jt{Phys. Rev. Fluids}  \bvol{5},  \pg{064802}.

\bibitem[Taylor(1922)]{taylor22}
{\sc \au{Taylor, G.~I.}} \yr{1922}  \at{Diffusion by continuous movements}.
  \jt{Proc. Lond. Math. Soc.}  \bvol{20},  \pg{196--212}.

\bibitem[Taylor(1938)]{taylor38}
{\sc \au{Taylor, G.~I.}} \yr{1938}  \at{Production and dissipation of vorticity
  in a turbulent fluid}.  \jt{Proc. Roy. Soc. Lond. A Mat.}  \bvol{164}~(916),
  \pg{15--23}.

\bibitem[Taylor {\em et~al.\/}(2016)Taylor, Deusebio, Caulfield \&
  Kerswell]{taylor16}
{\sc \au{Taylor, J.~R.}, \au{Deusebio, E.}, \au{Caulfield, C.~P.} \&
  \au{Kerswell, R.~R.}} \yr{2016}  \at{A new method for isolating turbulent
  states in transitional stratified plane couette flow}.  \jt{Journal of Fluid
  Mechanics}  \bvol{808},  \pg{R1}.

\bibitem[Tennekes \& Lumley(1972)]{tennekes}
{\sc \au{Tennekes, H.} \& \au{Lumley, J.~L.}} \yr{1972} {\em A first course in
  turbulence\/}.  \publ{Cambridge: MIT Press}.

\bibitem[Tom {\em et~al.\/}(2020)Tom, Carbone \& Bragg]{tom20}
{\sc \au{Tom, Josin}, \au{Carbone, Maurizio} \& \au{Bragg, Andrew~D}} \yr{2020}
  Exploring the turbulent velocity gradients at different scales from the
  perspective of the strain-rate eigenframe,  \arxiv{arXiv: 2005.04300}.

\bibitem[Towns {\em et~al.\/}(2014)Towns, Cockerill, Dahan, Foster, Gaither,
  Grimshaw, Hazlewood, Lathrop, Lifka, Peterson, Roskies, Scott \&
  Wilkins-Diehr]{xsede}
{\sc \au{Towns, J.}, \au{Cockerill, T.}, \au{Dahan, M.}, \au{Foster, I.},
  \au{Gaither, K.}, \au{Grimshaw, A.}, \au{Hazlewood, V.}, \au{Lathrop, S.},
  \au{Lifka, D.}, \au{Peterson, G.~D.}, \au{Roskies, R.}, \au{Scott, J.~R.} \&
  \au{Wilkins-Diehr, N.}} \yr{2014}  \at{Xsede: Accelerating scientific
  discovery}.  \jt{Computing in Science \& Engineering}  \bvol{16}~(5),
  \pg{62--74}.

\bibitem[Tsinober(2001)]{tsinober}
{\sc \au{Tsinober, Arkady}} \yr{2001} {\em An informal introduction to
  turbulence\/}.  \publ{Kluwer Academic Publishers}.

\bibitem[Vallis(2006)]{Vallis06}
{\sc \au{Vallis, G.~K.}} \yr{2006} {\em Atmospheric and Oceanic Fluid
  Dynamics\/}.  \publ{Cambridge, U.K.: Cambridge University Press}.

\bibitem[Vreman {\em et~al.\/}(1994)Vreman, Geurts \& Kuerten]{vreman94}
{\sc \au{Vreman, Bert}, \au{Geurts, Bernard} \& \au{Kuerten, Hans}} \yr{1994}
  \at{Realizability conditions for the turbulent stress tensor in large-eddy
  simulation}.  \jt{Journal of Fluid Mechanics}  \bvol{278},  \pg{351–362}.

\bibitem[Waite \& Bartello(2006)]{waite06}
{\sc \au{Waite, Michael~L.} \& \au{Bartello, Peter}} \yr{2006}  \at{The
  transition from geostrophic to stratified turbulence}.  \jt{Journal of Fluid
  Mechanics}  \bvol{568},  \pg{89?108}.

\bibitem[Wyngaard(2010)]{wyngaard}
{\sc \au{Wyngaard, J.~C.}} \yr{2010} {\em Turbulence in the Atmosphere\/}.
  \publ{Cambridge: Cambridge University Press}.

\bibitem[Zorzetto {\em et~al.\/}(2018)Zorzetto, Bragg \& Katul]{zorzetto18}
{\sc \au{Zorzetto, E.}, \au{Bragg, A.~D.} \& \au{Katul, G.}} \yr{2018}
  \at{Extremes, intermittency, and time directionality of atmospheric
  turbulence at the crossover from production to inertial scales}.  \jt{Phys.
  Rev. Fluids}  \bvol{3},  \pg{094604}.

\end{thebibliography}

\end{document}